\documentclass[epj]{svjour}
\usepackage{amssymb}
\usepackage{url}
\usepackage{graphics}
\usepackage[normalem]{ulem}   
\usepackage{color}  
\usepackage{graphics}
\usepackage{graphicx}
\usepackage{amssymb}
\usepackage{amsfonts}
\usepackage{amsmath}
\usepackage{amsbsy}
\usepackage[numbers,sort&compress]{natbib}
\usepackage{comment}
\usepackage{color}
\usepackage{subfigure}
\newcommand{\PHDG}{^{\phantom{\dagger}}}
\usepackage{graphicx}
\usepackage{textcomp}
\usepackage{amsmath,amssymb}
\usepackage{color}
\usepackage{romannum}
\usepackage{array}
\usepackage{xcolor}
\usepackage{soul}
\usepackage{mathtools}
\usepackage[colorlinks=true, urlcolor=blue, citecolor=blue, linkcolor=blue]{hyperref}
\newcommand{\cl}{\ell}
\newcommand{\tr}{\mbox{Tr}}
\begin{document}
\title{ Diagnosing chaos in a periodically driven Ising model with a ramping field via out-of-time-order correlation saturation}
\author{Rohit Kumar Shukla,\inst{1,2,3}\thanks{\email{rohitkrshukla.rs.phy17@itbhu.ac.in}} Gaurav Rudra Malik,\inst{4}\thanks{\email{gauravrudramalik.rs.phy22@itbhu.ac.in}} S. Aravinda,\inst{5}\thanks{\email{aravinda@iittp.ac.in}} Sunil Kumar Mishra\inst{4}\thanks{\email{sunilkm.app@iitbhu.ac.in}}}
\institute{Department of Chemistry; Institute of Nanotechnology and Advanced Materials; Center for
Quantum Entanglement Science and Technology, Bar-Ilan University, Ramat-Gan, 5290002, Israel \and Optics and Quantum Information Group, The Institute of Mathematical Sciences, CIT Campus, Taramani, Chennai 600113, India \and  Homi Bhabha National Institute, Training School Complex, Anushakti Nagar, Mumbai 400085, India \and Department of Physics, Indian Institute of Technology (Banaras Hindu University), Varanasi - 221005, India \and  Department of Physics, Indian Institute of Technology Tirupati, Tirupati, India 517619}
\date{\today}
\abstract{
The dynamic region of out-of-time-ordered correlators (OTOCs) serves as a powerful indicator of chaos in classical and semiclassical systems, capturing the characteristic exponential growth. In contrast, this signature fails to appear in spin systems, where even chaotic dynamics lack such exponential escalation, making this region an unreliable marker of chaos. To address this limitation, we turn to the saturation behavior of OTOCs to differentiate between chaotic and integrable regimes. In integrable systems, the saturation region of OTOCs exhibits oscillatory behavior, while in chaotic systems, it shows a stable saturation. To evaluate this distinction, we investigate a time-dependent Ising spin system subjected to a linearly ramping transverse field, analyzing both integrable (without longitudinal field) and non-integrable (with longitudinal field) scenarios. The ramping introduces a time-dependent increase of the external field, which influences the saturation regime of the OTOC—a region crucial for characterizing the chaotic behavior of the system. To quantify the degree of chaoticity, we compute the normalized Fourier spectrum of the OTOC and observe that increasing the ramping field strength leads to a suppression of oscillation frequencies in the saturation region of the OTOC, thereby enhancing the system's chaotic behaviour. To further support our findings, we investigate the level spacing distribution of time-dependent unitary operators, which effectively distinguishes chaotic from regular regions in our system and corroborates the results obtained from the saturation behavior of the OTOC. }

\maketitle
\section{Introduction}
\label{Introduction}
Studying quantum dynamics in many-body systems has been an exciting and challenging frontier in modern physics. Understanding the intricate behavior of quantum correlations, entanglement, and chaos across all dynamical regimes is essential for unlocking the full potential of quantum technologies and shedding light on fundamental aspects of quantum mechanics. 

 For the class of chaotic dynamics, pioneering research by Larkin and Ovchinnikov delved into the examination of out-of-time-order correlations (OTOCs) as a semiclassical approach to theoretically understand superconductivity\cite{larkin1969quasiclassical}. Their study relates the sensitivity of classical trajectories towards the initial conditions to an exponential growth of OTOC. Since then, the concept of OTOC has emerged as a powerful tool to explore the scrambling of information, quantum chaos, butterfly effect  \cite{prakash2020scrambling,rozenbaum2017lyapunov,aleiner2016microscopic,roberts2016lieb, huang2017out,wei2018exploring,yao2016interferometric}, and phase transition \cite{heyl2018detecting,chen2020detecting, shukla2021}, providing information about the chaotic nature of the classical and semiclassical systems \cite{prakash2020scrambling,maldacena2016bound}. A notable development in this field is the recent proposal of a quantum information diode that harnesses OTOC \cite{shukla2023quantum}. The exponential growth of OTOC represents the chaotic nature of the systems with a classical analogue, with its exponent being proportional to the Lyapunov exponent\cite{rozenbaum2017lyapunov}. However, it is not conversely true as despite the exponential growth of OTOC, the system in question may be integrable \cite{lin2018out}. Recently, researchers have utilized the norm of the adiabatic gauge potential to regulate the adiabatic transitions between eigenstates, revealing its effectiveness as a remarkably sensitive indicator of quantum chaos \cite{pandey2020adiabatic}. The quasiprobability distribution of the OTOC is also used to distinguish between integrable and scrambled systems.\cite{gonzalez2019out}.
\par
The growth of OTOC's has been extensively explored in various spin systems, including Luttinger liquids \cite{dora2017out}, the XY model \cite{bao2020out}, the Sachdev-Ye-Kitaev (SYK) model \cite{Fu2016}, the Heisenberg model, the XXZ model \cite{Riddell2019, Lee2019typical}, integrable transverse field Ising model (TFIM) \cite{Lin2018}, integrable and nonintegrable Floquet system \cite{shukla2022out}. Numerous experimental approaches for measuring OTOC have been proposed in the literature, spanning a wide range of platforms such as cold atoms, cavity and circuit quantum electrodynamics (QED), and trapped-ion simulations \cite{yao2016interferometric,swingle2016measuring,zhu2016measurement,campisi2017thermodynamics}. These ideas have paved the way for successful experimental realizations using various systems, such as nuclear spins of molecules \cite{wei2018exploring,chen2020detecting}, trapped ions \cite{landsman2019verified,joshi2020quantum}, and ultra-cold gases \cite{meier2019exploring}.
It is essential to highlight that the observed behavior consistently demonstrates the power-law growth of the OTOCs in both integrable and chaotic spin systems \cite{shukla2022characteristic,shukla2022out}. This implies that the dynamic region of OTOC does not allow for the differentiation between integrable and chaotic spin systems, as it is limited by the finite-dimensional Hilbert space.
\par
The OTOC is frequently discussed in reference to the Floquet spin Hamiltonian driven by time-periodic fields \cite{shukla2022out,shukla2022characteristic,kukuljan2017weak,shukla2025out}. The dynamic and saturation characteristics of  OTOCs using the nonlocal spin block observables for the Floquet Ising model subjected to constant magnetic fields, encompassing both integrable and nonintegrable cases, have been studied\cite{shukla2022out}. Notably, the observed behavior of OTOCs in both integrable and chaotic Floquet systems revealed an approximately quadratic power-law growth pattern. If we consider local spin observables, the OTOCs still exhibited a power-law growth in their dynamic region in both integrable and chaotic cases\cite{shukla2022characteristic}. Consequently, we can say that OTOC in constant fields Floquet system calculated by using different types of observables like single spin observables \cite{shukla2022characteristic}, spin block observables \cite{shukla2022out}, and extensive sums of local observables \cite{kukuljan2017weak}, does not provide exponential growth of OTOC irrespective of a chaotic system. 
\par
In this work, we explore the behavior of OTOC in a time-dependent Ising spin system under a periodically kicked transverse field, where the field strength increases over time. This setup accelerates the system’s randomization, presumably enhancing chaotic dynamics in the non-integrable regime and making the model particularly compelling for our study. We examine both the dynamic and saturation regions of the OTOC, uncovering distinct saturation behaviors that differentiate integrable and chaotic systems.
\par
This paper is structured as follows. We begin by introducing a time-dependent Ising spin system. Next, we define OTOCs and the associated spin observables. We then discuss the level spacing statistics (LSS) within the framework of the system under consideration. Following this, we present our findings. Finally, we conclude by summarizing our main contributions.
\maketitle
\section{Theoretical Setup and Diagnostic Tools}
\subsection{Time-dependent Ising spin system}
\label{model}
We consider an Ising chain of length $N$ with open boundary conditions, where the quantum spins are described by Pauli matrices $\sigma_l^x$, $\sigma_l^y$, and $\sigma_l^z$ at site $l$. We apply a time-dependent transverse magnetic field with a magnitude $h_z$  that varies like a straight line with slope $\Gamma$ and intercept $h_{z0}$. In the mathematical form, $h_z$ can be represented as $h_z(t)=h_{z0}+\Gamma t$. This field is applied in the form of kicks (represented as integer $n$) with period $\tau$.  A small constant longitudinal magnetic field of amplitude $h_{x0}$ is also applied in order to make the system nonintegrable. 
\par
The quantum map represents the time evolution operator that evolves the system from \( n\tau^+ \) to \( (n+1)\tau^+ \), where \( \tau^+ \) denotes the time just after \( \tau \). The quantum map for a time-dependent Ising spin system is denoted by \( \mathcal{\hat{U}}_{xl}(n) \) and is defined as follows (hereafter referred to as the \( \mathcal{\hat{U}}_{xl} \) system):
\begin{eqnarray}
\label{Uxq}
\mathcal{\hat U}_{xl}(n)=&&\Big[\prod_{l=1}^{N}\exp \Big(-i \tau J \hat \sigma_l^x \hat \sigma_{l+1}^x - i\tau h_x  \hat \sigma_l^x \Big) \Big] \nonumber \\
&&\Big[\prod_{l=1}^{N}\exp(-i\tau h_z(n\tau) \hat \sigma_l^z)\Big].
\end{eqnarray}
Here, $J$ represents the interaction strength between neighbouring spins, and $h_x/h_z$ ($h_x=h_{x0}$) is the longitudinal/transverse field. The time evolution of operator $\hat W(0)$ after $n$ kicks is defined as $\hat W(n)=\hat U^{\dagger}(n)\hat W(0)\hat U(n)$, where  $\hat U(n)$ is time evolution operator and defined as $\hat U(n)= \mathcal{\hat U}_{xl}(1) \mathcal{\hat U}_{xl}(2)\mathcal{\hat U}_{xl}(3) \cdots \mathcal{\hat U}_{xl}(n)$. The amplitude of the transverse field varies with time, rendering the model non-periodic in nature {\it i.e.} $\hat U(t+n\tau)\neq \hat U(t)$. 
 \par
\subsection{ Out-of-time-order correlation}
\label{OTOC}
Consider two Hermitian observables $\hat W(0)$ and $\hat V(0)$ at time $t=0$. After the time evolution of observable $\hat W(0)$ {\it i.e.} $\hat W(t=n\tau)$, it displays noncommutativity with $\hat V(0)$ which provides the dynamics of the quantum systems by the spreading of operators under the system's unitary time evolution. This noncommutativity may capture chaotic behavior in the system, which is widely known as OTOC. In mathematical form, OTOC is defined as \cite{larkin1969quasiclassical,Hashimoto2017}
\begin{eqnarray}
C(n)=-\frac{1}{2}\mbox{Tr} \left(\hat \rho[\hat W(n), \hat V]^2\right),
\label{Cn1}
\end{eqnarray}
where $\hat{\rho}$ is the state of the system, defined as $\hat{\rho} = \frac{e^{-\beta \hat{H}}}{\text{Tr}[e^{-\beta \hat{H}}]}$, with $\beta = \frac{1}{k_B T}$, $k_B$ is the Boltzmann constant and $T$ is the temperature of the system. For our calculation of OTOC, we consider the infinite temperature limit, in which case $\hat{\rho} = \frac{1}{2^N}$. We focus on nonlocalized observables, defined as follows:


 

Consider a chain of length \( N \) (with \( N \) being even) divided into two equal halves. We define an observable \( \hat{W} \) over the first half and an observable \( \hat{V} \) over the second half. Mathematically, these are given by:
$
\hat{W} = \frac{2}{N} \sum_{l=1}^{\frac{N}{2}} \hat{\sigma}_l^x \quad \text{and} \quad \hat{V} = \frac{2}{N} \sum_{l=\frac{N}{2}+1}^{N} \hat{\sigma}_l^x.
$
Note that the observables \( \hat{W} \) and \( \hat{V} \) are \textit{Hermitian but nonunitary}. For such observables, the OTOC, as defined by Eq. (\ref{Cn1}), can be simplified as follows:

\begin{align}
C(n)=\frac{1}{ 2^N}\mbox{Tr} \left(\hat W^2(n) \hat V^2\right) - \frac{1}{ 2^N}\mbox{Tr} \left(\hat W(n) \hat V\hat W(n) \hat V\right).
\label{c2c4}
\end{align}
In this expression, the first term is called as the two-point correlation, and the second term is the four-point correlation. For \( n > 0 \), \( \hat{W}(n) \) is no longer confined to the first \( N/2 \) spins, and the OTOC becomes nonzero. Since our observable is non-unitary, the two-point correlation does not reduce to unity. As \( n \rightarrow \infty \), it asymptotically approaches \( \frac{4}{N^2} \), while the four-point correlation tends to zero. Thus, in the long-time limit, \( C(\infty) = \frac{4}{N^2} \). The normalized value of the OTOC, \( C_N(n) \), relative to its long-time limit is given by
$
C_N(n) = \frac{C(n)}{C(\infty)}.
$
\par
The motivation for considering non-local operators is to increase the Hilbert space dimension of the observables compared to single-site observables. For such observables, information propagation occurs instantly, unlike in localized observables at different positions, allowing for more rapid saturation. Since our focus is on the saturation behavior of the OTOC, we aim to reach the saturation region of the OTOC more quickly.

\subsection{Level Spacing Statistics}
\label{NNSD}
The level spacing statistics (LSS) is a concept commonly used in the field of  chaos and random matrix theory to study the statistical properties of energy levels of complex systems \cite{mehta1991theory,santos2004integrability,bogomolny1992chaotic,bohigas1984characterization}. It provides insights into the level repulsion and clustering behavior of eigenvalues in such systems. The LSS is therefore particularly relevant when studying systems with chaotic and regular dynamics as it  characterizes the distribution of spacings between adjacent energy levels, distinguishing a chaotic system from an integrable one \cite{alhassid1989nearest,prosen1993energy}. 
 \par
In spin systems with constant longitudinal and transverse fields, the Hamiltonian can be used to describe the level statistics of the system. However, when the field is applied in the form of periodic kicks, the Hamiltonian includes a Dirac delta function term, classifying the system as a Floquet system. This delta function term makes it challenging to use the Hamiltonian directly for analyzing the level statistics \cite{szasz2021weak}.

In such cases, an alternative approach is employed, which involves the spectral analysis of the unitary operator governing the system’s evolution over one period. In Floquet systems, the unitary operator exhibits periodic behavior, {\it i.e.,} $\mathcal{\hat U}_{xl}( t +n\tau)=\mathcal{\hat U}_{xl}(t)$. The spectral properties of the system can be fully characterized by the unitary operator for a single period. By taking the logarithm of the eigenvalues of the unitary operator, we obtain the eigenvalues of the system’s Hamiltonian, and the level statistics of these eigenvalues represent the statistical behavior of the system \cite{PhysRevE_Floquet,PhysRevX_Floquet}.
\par
If the amplitude of the kicking field varies with time, the evolution becomes aperiodic, meaning $\mathcal{\hat U}_{xl}(t + n\tau) \neq \mathcal{\hat U}_{xl}(t)$. In this case, the unitary operators for different periods are distinct, and the system’s evolution depends on the specific time ordering of the kicks. Consequently, we cannot describe the spectral statistics of the system using the unitary operator for just a single period, as we would in a Floquet system. To accurately describe the spectral statistics for such a system, we must consider the unitary operators for all individual periods. For example, if we want to find the spectral statistics at first, second, and third kicks/period, we would consider the following unitary operators $\hat U(1)=\mathcal{\hat U}_{xl}(1)$, $\hat U(2)=\mathcal{\hat U}_{xl}(1)\mathcal{\hat U}_{xl}(2)$, and $\hat U(3)=\mathcal{\hat U}_{xl}(1)\mathcal{\hat U}_{xl}(2)\mathcal{\hat U}_{xl}(3)$, respectively. We would then calculate the spectral statistics of each of these unitary operators individually to describe the spectral statistics at these three different kicks. If we are interested in the spectral statistics over a larger number of kicks, we would follow a similar process by considering the corresponding unitary operators for each kick.
\begin{figure}
 \includegraphics[width=.95\linewidth, height=.60\linewidth]{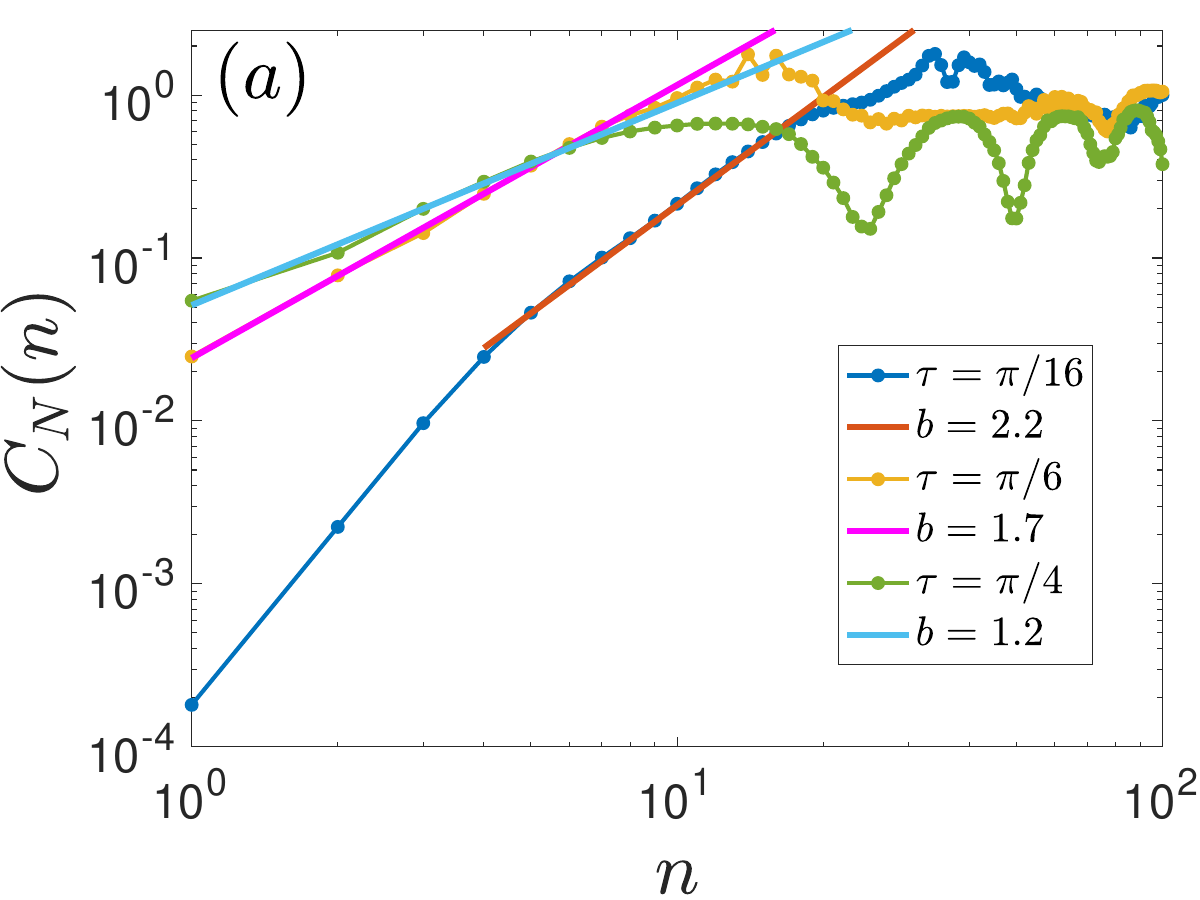} 
  \includegraphics[width=.95\linewidth, height=.60\linewidth]{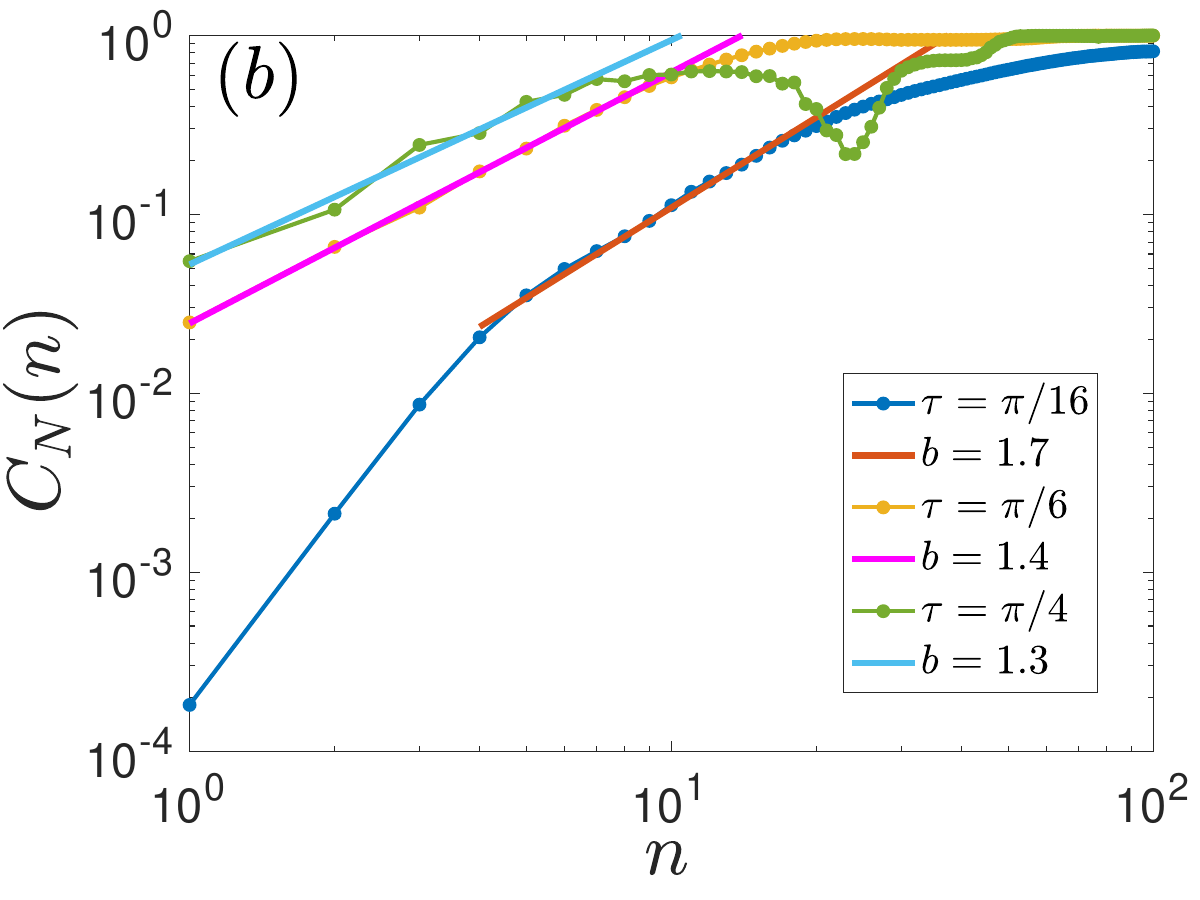} 
 \caption{$C_N(n)$ vs $n$ for ($a$) integrable $\mathcal{\hat U}_{0l}$ ($h_{x0}=0$) ($b$) nonintegrable $\mathcal{\hat U}_{xl}$ system ($h_{x0}=1$) for period $\tau=\pi/16,~\pi/6$, and $\pi/4$.   The parameters used for the numerics are $J=1$, $h_{z0}=1$, $\Gamma=0.1$ and $N=12$. $b$ is the power exponent   {\it i.e., }$C_N(n)\sim n^b$. The open boundary condition is considered.}
\label{linear_growth} 
\end{figure}

\begin{figure*}
 \includegraphics[width=.32\linewidth, height=.25\linewidth]{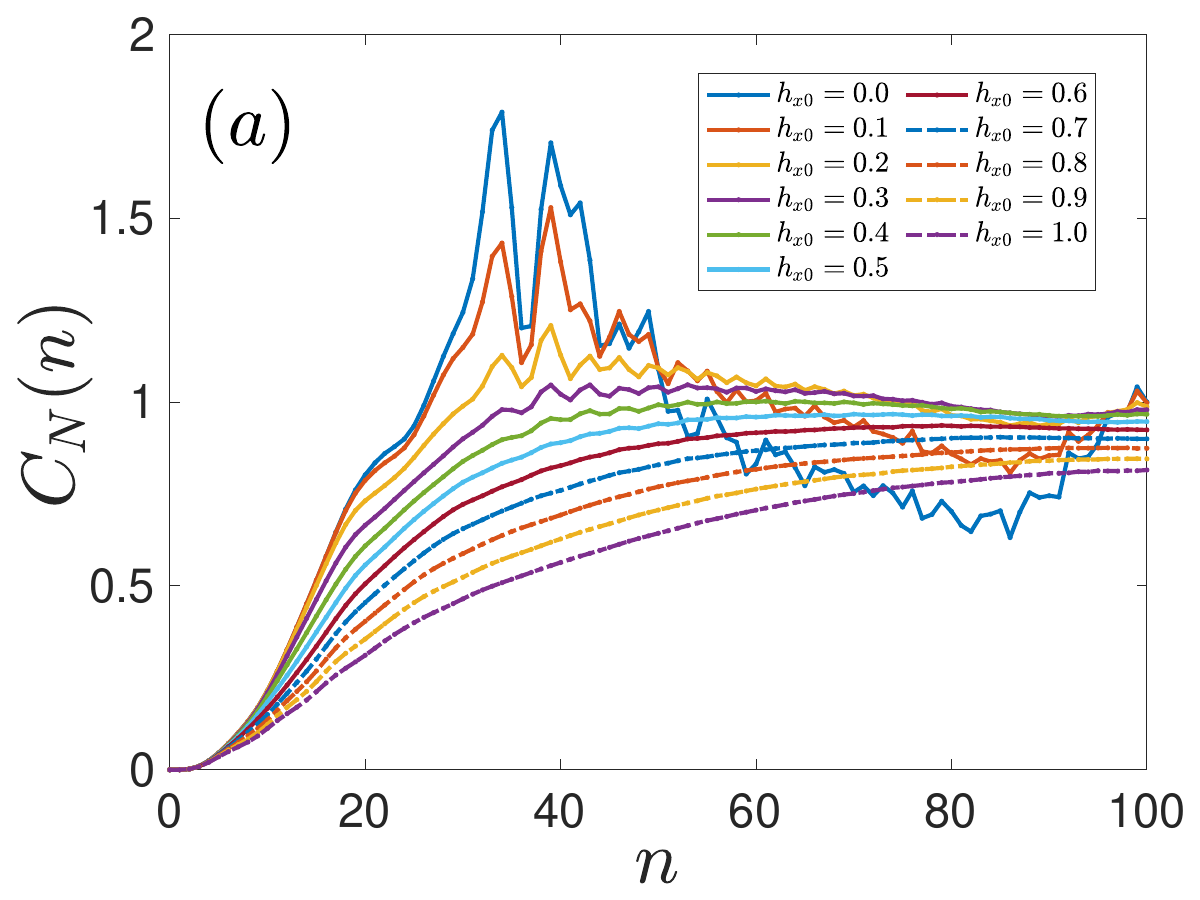} 
 \includegraphics[width=.32\linewidth, height=.25\linewidth]{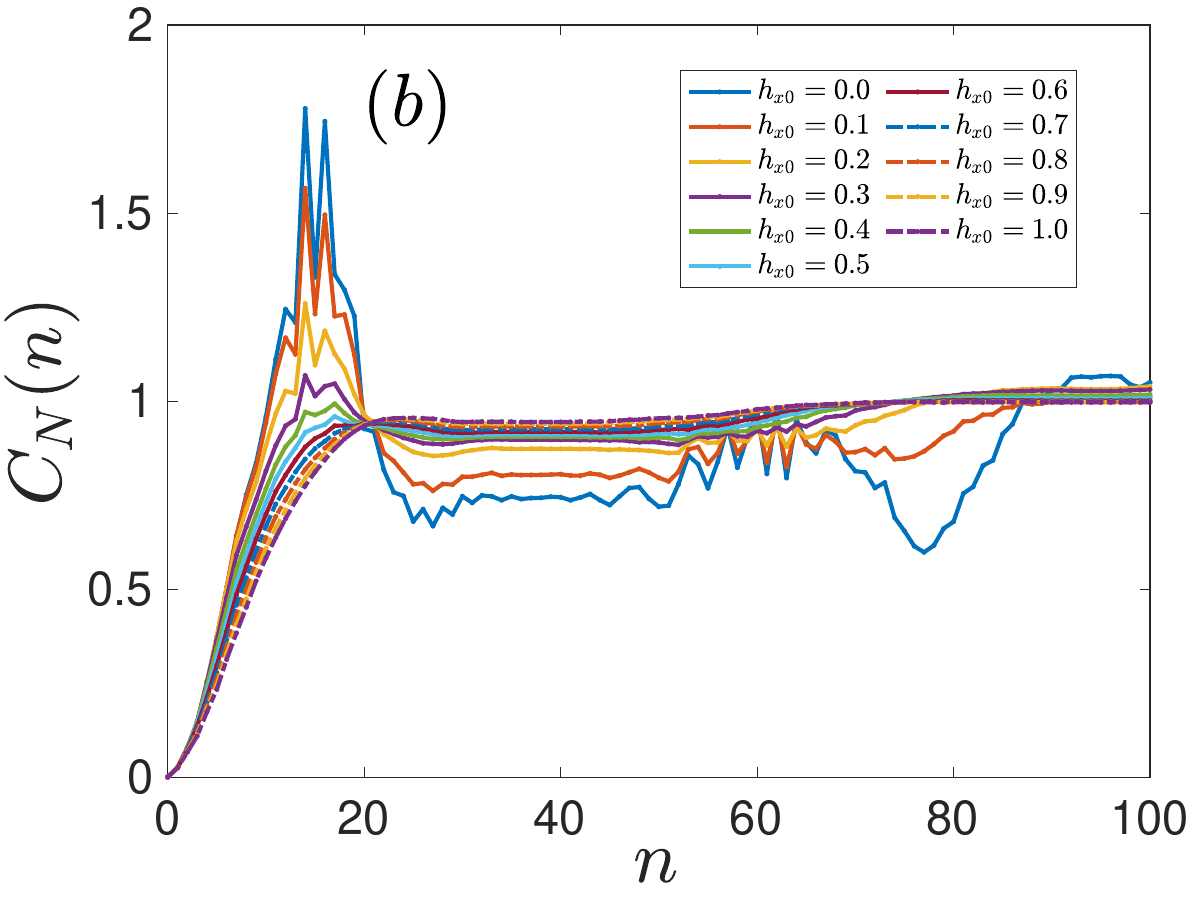}
  \includegraphics[width=.32\linewidth, height=.25\linewidth]{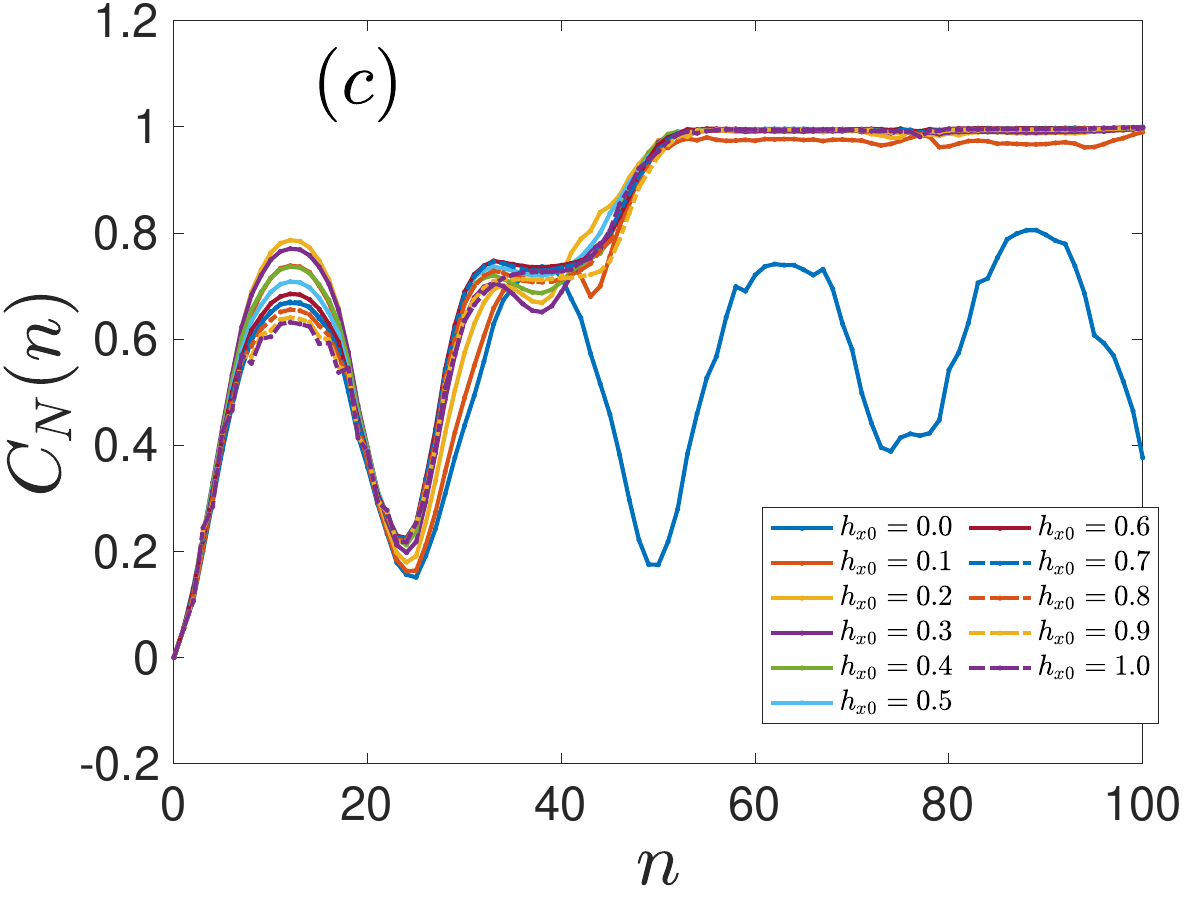}
 \caption{$C_N(n)$ vs $n$ for different value of  $h_{x0}$  lies in between $0$ to $1$ differing by $0.1$ for nonintegrable $\mathcal{\hat U}_{xl}$  system at period $(a)$ $\tau=\pi/16$, $(b)$ $\tau=\pi/6$, and $(c)$ $\tau=\pi/4$. Other parameters are: $J=1$, $h_{z0}=1$, $\Gamma=0.1$, and $N=12$ . Open boundary conditions are considered.}
\label{linear_sat} 
\end{figure*}
\par
\begin{figure}
 \includegraphics[width=.95\linewidth, height=.60\linewidth]{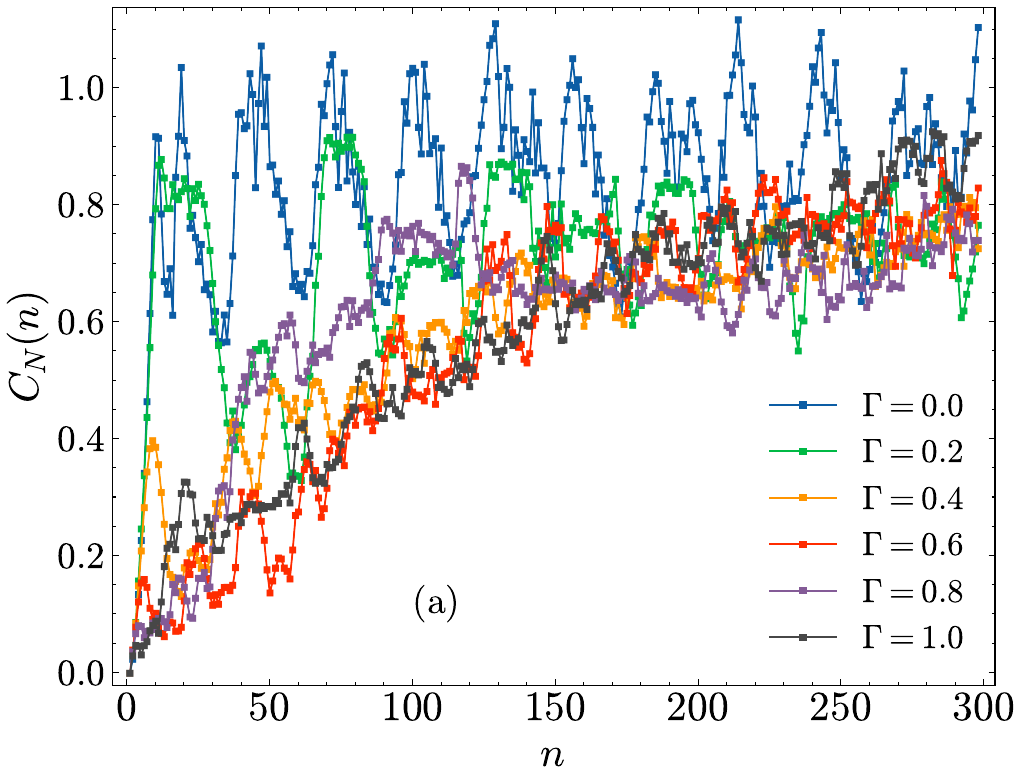} 
  \includegraphics[width=.95\linewidth, height=.60\linewidth]{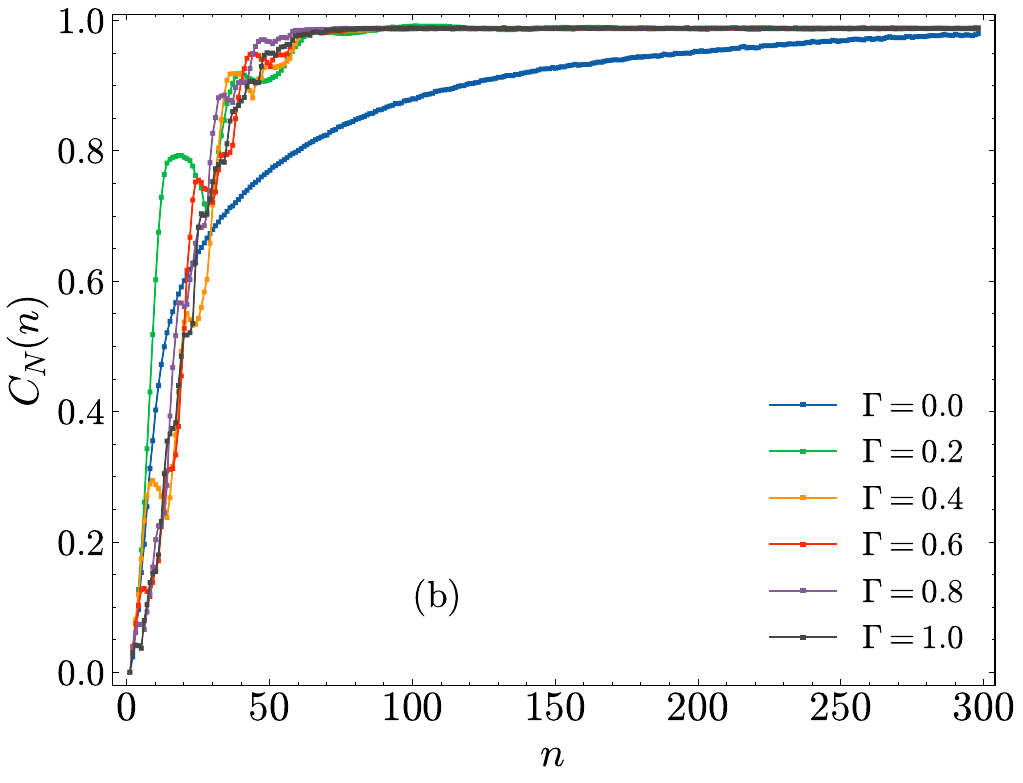}
 \caption{ $C_N(n)$ vs $n$ for $(a)$ integrable $\mathcal{\hat U}_{0l}$ and $(b)$ nonintegrable $\mathcal{\hat U}_{xl}$ system with period $\tau=\pi/6$ at different value of $\Gamma$.  Other parameters are: $J=1$, $h_{x0}=0/1$, $h_{z0}=1$, and $N=12$. Open boundary conditions are considered.}
\label{linear_gamma} 
\end{figure}
\begin{figure}
\includegraphics[width=\linewidth, height=.60\linewidth]{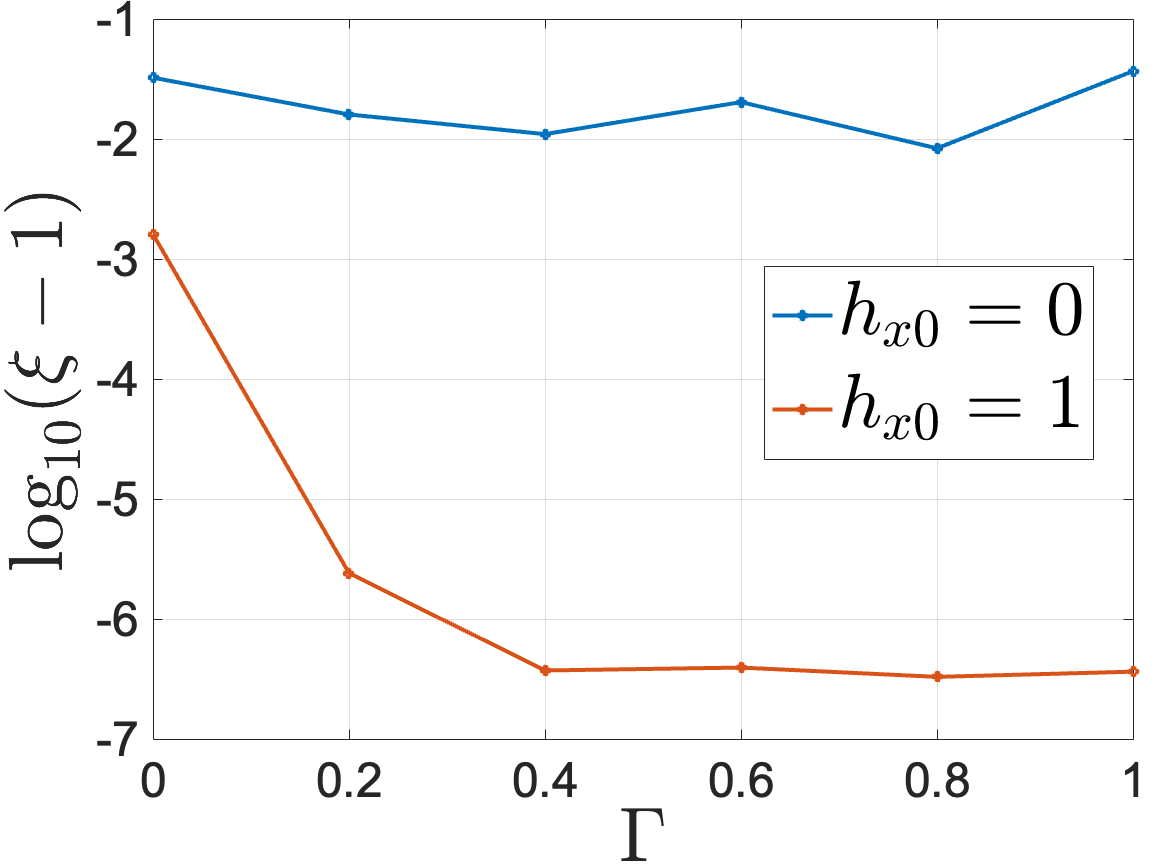} 
\caption{Fourier space analysis of the OTOC, shown in Fig. \ref{linear_gamma} (a, b), presented as the variation of $\log_{10}(\xi - 1)$ with respect to $\Gamma$ for a fixed maximum time step $n_{\rm max} = 300$.}
\label{fourier_figures}
\end{figure}

\par
To accurately calculate the LSS, it is essential to eliminate system symmetries that contribute to degeneracies. In a time-dependent system with a periodic external field and open boundary conditions, reflection symmetry is present. This symmetry can be removed by dividing the eigenvector space into  either even and odd palindromic subspaces. We generate the Hamiltonian taking the even case for our calculations (for details, see Ref. \cite{shukla2022out}) and then compute the differences between consecutive eigenvalues of the Hamiltonian within  the selected subspaces, with these differences forming ensembles denoted as $s$. After obtaining the ensemble of $s$, we can draw the distribution of these ensembles. This distribution is known as LSS. If the LSS closely adheres to the Wigner-Dyson (WD) distribution, the system exhibits strong chaotic behavior. The WD distribution is mathematically expressed as \cite{Luca2016,mehta1991theory}:  $P_W(s)=\frac{\pi s}{2}e^{-\pi s^2/4}$. Conversely, if the LSS exhibits Poisson type, the system is termed nontrivially integrable. The Poisson distribution is given by: $P_P(s)=e^{-s}$.

\section{Results}
\label{result}
We investigate the OTOC within time-dependent Ising spin systems, considering both integrable and non-integrable cases. To calculate the OTOC, we employ non-local observables.  OTOC display approximately quadratic power-law growth in both integrable [Fig.~\ref{linear_growth}($a$)] and nonintegrable systems [Fig.~\ref{linear_growth}($b$)] for a small period {\it i.e,} $C_N(n) \sim n^b$ where $b\approx 2$, $b$ is known as the power exponent.  As the period of the transverse kicks increases, we observe a decrease in the associated exponent of the power law growth. This contrasts with the behavior of the  nonintegrable system under a constant field, where a longer period leads to an increase in the power law exponent \cite{shukla2022out}. In the current scenario, there is a time-dependent transverse magnetic field. As the period increases, the transverse kick amplitude becomes rapidly dominant due to its magnitude being defined by $h_z(n\tau) = h_{\rm z0} + n \tau \Gamma$ and saturates in a shorter number of kicks. This in turn leads to a suppression in the growth exponent.
\par
\begin{figure*}
 \includegraphics[width=.49\linewidth, height=.30\linewidth]{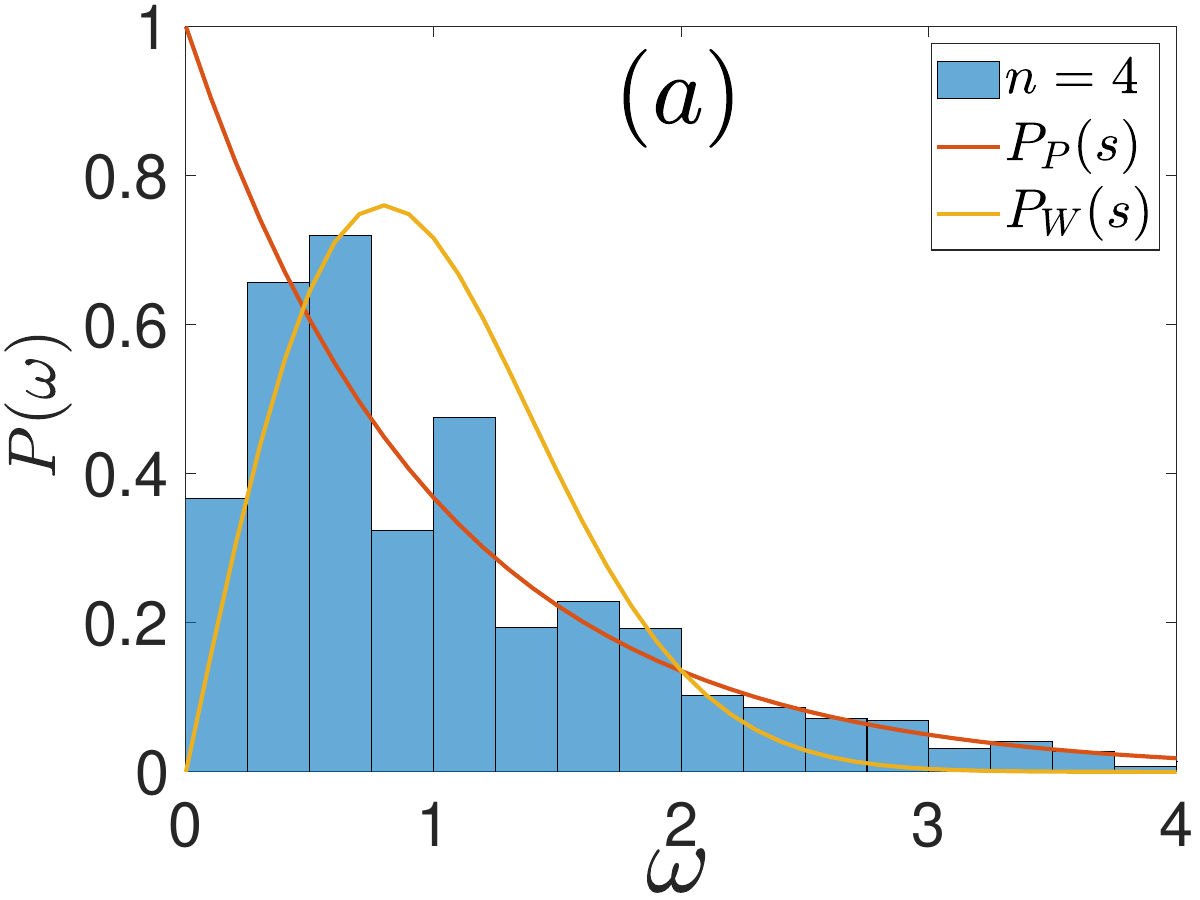}
     \includegraphics[width=.49\linewidth, height=.30\linewidth]{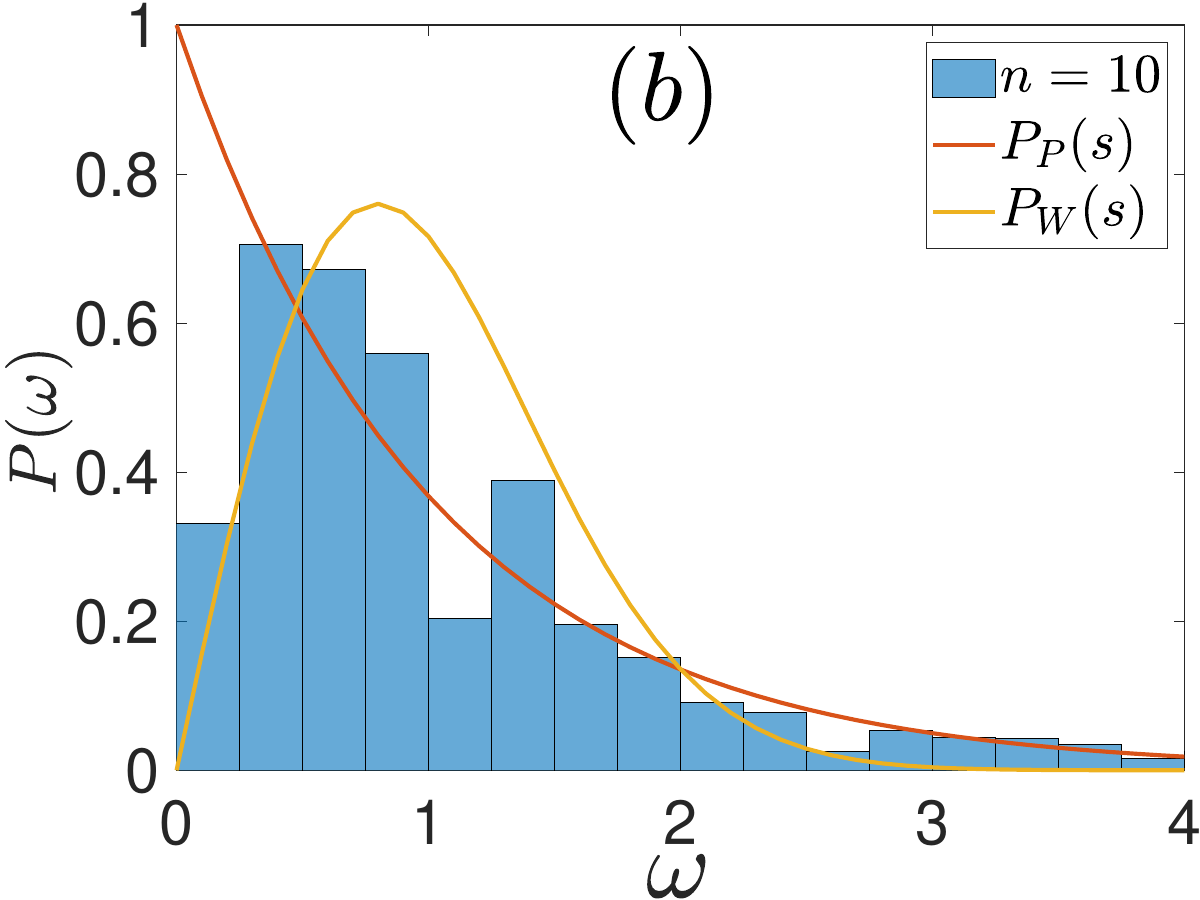}
  \includegraphics[width=.49\linewidth, height=.30\linewidth]{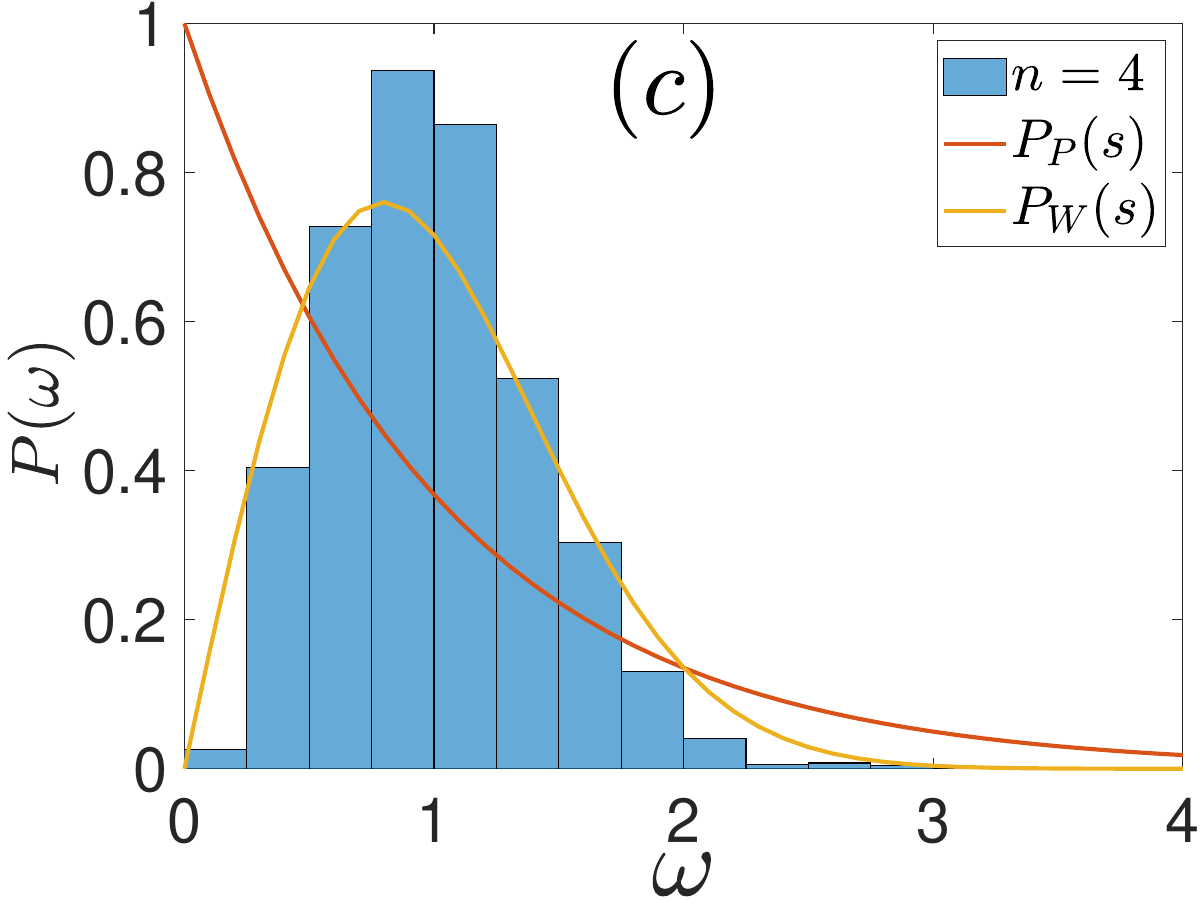}
    \includegraphics[width=.49\linewidth, height=.30\linewidth]{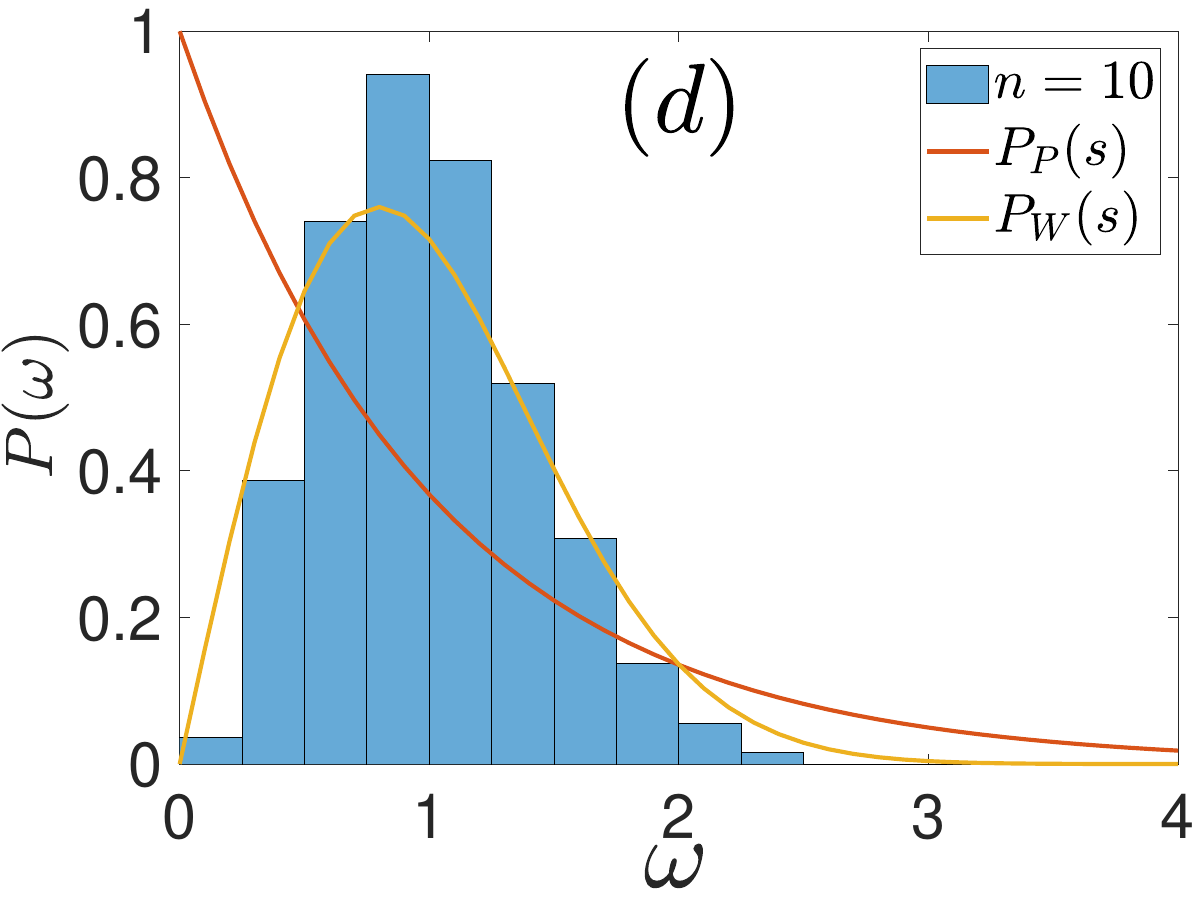}
  \caption{LSS of the integrable $\mathcal{\hat U}_{0l}$ system ($a$ and $b$), and nonintegrable $\mathcal{\hat U}_{xl}$ system ($c$ and $d$) at period $\tau=\pi/4$ in the even Palindromic space at different kicks mentioned in the legend.  Other parameters are: $J=1$, $h_{z0}=1$, $h_{x0}=0/1$, $\Gamma=0.1$, and $N=12$. Open boundary conditions are considered. }
\label{linear_NNSD}
 \end{figure*}
Since the dynamic region of the OTOC does not effectively distinguish between chaotic and regular dynamics in spin systems, we shift our focus from early-time growth to the saturation behavior of the OTOC, which offers a more promising indicator of underlying chaoticity. To this end, we analyze the saturation behavior of the OTOC in a time-dependent Ising spin chain. Our investigation considers three distinct values of the driving period: $\tau = \pi/16$, $\tau = \pi/6$, and $\tau = \pi/4$. These values are chosen to capture a range of dynamical regimes—$\tau = \pi/16$ corresponds to a relatively short period, $\tau = \pi/6$ represents a longer driving cycle, and $\tau = \pi/4$ serves as a special point known to exhibit periodic revivals in the Floquet dynamics, as previously demonstrated in Refs.~\cite{shukla2022out, naik2019controlled}. However, in the time-dependent case, it does not exhibit periodic behavior due to the constant ($h_x/h_z$) being multiplied with the period $\pi/4$, unlike the special case where $h_x=h_z=1$. In our exploration of the saturation behavior of OTOC, we set the longitudinal field strength from $0$ to $1$ in increments of $0.1$. This approach helps us understand the effect of the longitudinal field in the saturation region of the OTOC. The presence of the longitudinal field makes the system nonintegrable. When the longitudinal field is absent, oscillations are present in the OTOC behaviour for all periods. However, as we introduce a small longitudinal field term, the oscillation amplitude decreases, and after a certain value of $h_{x0}$, it saturates at a specific value [Fig.~\ref{linear_sat} ($a-c$)]. At the period $\tau=\pi/4$, we observe exact saturation in the long-time regime of the OTOC with the introduction of a small longitudinal field term [Fig. \ref{linear_sat} ($c$)]. The exact saturation of the OTOC indicates the chaotic behavior of its underlying dynamics. Hence, we may say that a specific value of the longitudinal field term drives the system into chaos.
\par 
 We also examine how the growth rate of the transverse field, \( \Gamma \), influences the saturation behavior of the OTOC. In the integrable regime, the saturation region consistently exhibits oscillations for all values of \( \Gamma \), whereas in the non-integrable regime, the OTOC tends toward a steady saturation, as illustrated in Fig.~\ref{linear_gamma}($a$,$b$). Interestingly, the OTOC shows a similar saturation response when varying the kick period \( \tau \), highlighting the complementary relationship between \( \tau \) and \( \Gamma \), given by the relation \( h_z(t) = h_{z0} + \Gamma n \tau \). 
 \par

In the chaotic system, when the ramping field is absent ($\Gamma = 0$), the transverse field remains constant, $h_z(t) = h_{z0}$, effectively reducing the system to a conventional Floquet model with a static field. In this case, the OTOC grows gradually and reaches saturation only after a long time. However, the introduction of a ramping field results in a notable change in the dynamics: the OTOC rises more rapidly and saturates over a much shorter timescale, indicating increased response to the applied ramping field $\Gamma \neq 0$ compared to the static field case ($\Gamma = 0$). Remarkably, the OTOC curves for different ramping strengths, ranging from $\Gamma = 0.2$ to $1$, collapse onto a single trajectory, as shown in Fig.~\ref{linear_gamma}($b$). This distinct behavior highlights the nontrivial role of the ramping field in enhancing chaotic characteristics of the system, setting it apart from the response observed in conventional Floquet models with time-independent fields. Additionally, oscillations with multiple frequency components appear in the saturation region of the OTOC, and their characteristic frequencies decrease as the ramping strength increases. However, these oscillations are not straightforwardly understood from the time-domain behavior of the OTOC alone. To explore their origin and structure more thoroughly, we perform a Fourier space analysis of the OTOC in the following section.

\par
{\it Fourier Space Analysis:} The saturation behavior of OTOCs fail to fully capture the influence of the ramping field strength $\Gamma$ in order to describe the order of chaoticity of the system. To overcome this limitation, we perform a Fourier space analysis of the OTOC time series, which serves as a more sensitive probe of the underlying dynamical features. Specifically, we examine the time series $\{ C_N(n) \}_{n = 1}^{n_{\rm max}}$, representing OTOC values sampled at regular intervals $\tau$, where $n_{\rm max}$ is the total number of observations. A discrete Fourier transform of this series yields the frequency components $\{ f_j \}_{j = 1}^{n_{\rm max}}$, with $|f_j|$ denoting the amplitude of the frequency mode $2\pi j / n_{\rm max}$.
\par
To quantify the number of frequencies in the saturation region of OTOC, we normalize the Fourier components such that $\sum_j |f_j|^2 = 1$, and define the metric
\[
\xi = \left(\sum_j |f_j|^4\right)^{-1},
\]
which reflects the effective number of significant frequency components in the signal. For a time-independent or constan OTOC signal, only the zero-frequency (d.c.) mode contributes, yielding $\xi = 1$, the minimum possible value. Oscillatory signals involve multiple frequency modes and hence result in $\xi > 1$. The normalization ensures that $\sum_j |f_j|^4 \leq 1$, which implies $\xi \geq 1$; equality holds only when a single frequency component dominates. The manner in which $\xi$ is defined is reminiscent of the inverse participation ration (IPR) measure used in many-body physics to study localisation in physical systems. There a high (low) value of IPR indicates the spread of a wavefunction $|\psi\rangle$ over several (few) members of the eigenbasis.
\par
To quantify the effect of $\Gamma$ in the saturation regime, we evaluate $\log_{10}(\xi - 1)$—the logarithmic deviation from purely static dynamics. In the integrable regime ($h_x = 0$), this quantity remains between $-1$ and $-2$ across the full range $\Gamma \in [0, 1]$, indicating low but stable frequency complexity.  In contrast, in the chaotic regime ($h_x = 1$) with no ramping field ($\Gamma = 0$), the value of $\log_{10}(\xi - 1)$ is approximately $-3$, indicating the presence of nontrivial frequency components despite the chaotic nature of the system. As the ramping field is gradually introduced, $\log_{10}(\xi - 1)$ decreases steadily and eventually saturates near $-6.5$ beyond a critical strength of approximately $\Gamma = 0.4$ (see Fig.~\ref{fourier_figures}). This pronounced suppression of high-frequency components demonstrates that the ramping field progressively regularizes the chaotic dynamics, serving as a clear indicator of reduced frequency complexity in the saturation regime.

\par
{\it Level spacing statistics:} To verify the dynamics of our system, we calculate the LSS within the context of time-dependent unitary operators, introducing a novel dimension to spectral analysis. The lack of such an investigation in the existing literature highlights the originality of our approach. We compute the LSS for both integrable and non-integrable systems at the 4th and 10th kicks. In both cases, the integrable system exhibits a clustered LSS that follows a Poisson distribution, as shown in Fig.\ref{linear_NNSD}$(a-b)$. In contrast, for the non-integrable system, the LSS undergoes a marked transition to a Wigner-Dyson distribution, illustrated in Fig.\ref{linear_NNSD}$(c-d)$. This shift indicates level repulsion, which is characteristic of quantum chaos. This finding is particularly noteworthy, as the relationship between level spacing statistics and the distinction between integrable and chaotic dynamics has been extensively studied primarily in the context of Floquet systems. We observe that even after moving away from the Floquet regime, the results remain consistent, at least for our specific case \cite{PhysRevE_Floquet,PhysRevX_Floquet}.

\par

\section{Conclusion}
\label{conclusion}
We study OTOCs using non-localized observables for a time-dependent Ising spin system. Our focus is on understanding how these OTOCs evolve in the presence and absence of longitudinal fields for the non-integrable/integrable versions of the system, exploring their growth and saturation characteristics. 
\par
In both integrable and chaotic cases, we observe a consistent pattern of power-law growth in the OTOCs, characterized by a quadratic exponent over shorter periods that diminishes with longer driving periods. Interestingly, the dynamic region reveals no clear distinctions between integrable and chaotic systems, indicating that initial dynamics do not differentiate chaotic behavior. Instead, it is the saturation behavior of the OTOCs that serves as a reliable indicator of system properties: in the integrable system, OTOCs exhibit oscillatory behavior in the saturation regime, while in chaotic systems, they stabilize at a fixed value. This distinction offers a clear criterion for differentiating between the two regimes. 
\par
The oscillations observed in the saturation behavior of the OTOC are influenced by the growth rate of the transverse field but are not directly evident from the OTOC itself. To uncover this dependence, we perform a Fourier analysis of the OTOC in both integrable and non-integrable regimes, characterized by the absence and presence of a longitudinal field, respectively. In the integrable case (without the longitudinal field), oscillations persist in the saturation region of the OTOC across all values of $\Gamma$. In contrast, in the chaotic system, these oscillations gradually diminish as the ramping strength $\Gamma$ increases, indicating a suppression of frequency components due to enhanced chaoticity. 
\par
Validating our results enables us to propose a spectral measure for time-dependent systems. This additional analysis consistently reinforces our earlier observations, providing strong evidence for the distinction between integrable and chaotic systems. In the integrable case, the LSS exhibits Poisson-type behavior, whereas in the chaotic case, it displays Wigner-Dyson-type behavior.
\par
To generalize our results 
and explore the applicability of our discriminator, we also present findings for the case where both the longitudinal and transverse magnetic fields are time-dependent, represented as cosine and sine waves, respectively. In this scenario, we explore how different observables impact the saturation behavior of the OTOC in the systems with periodic driving fields. Specifically, we consider two types of observables. First, we analyze non-localized observables and localized single-spin observables where the spins are aligned in the $x$-direction. Corresponding to these scenarios, results are provided in SM-\Romannum{1}. Additionally, we consider both non-localized and localized observables oriented in the $z$-direction. Detailed discussion and the results can be found in SM-\Romannum{2}. By doing these analyses, we observe a consistent pattern of saturation behavior. In integrable systems, the OTOCs exhibit oscillatory behaviour, while in chaotic systems, they saturate at specific values.  
\par
 In conclusion, our findings suggest that the saturation behavior of the OTOC can serve as a robust indicator to distinguish between integrable and chaotic systems, irrespective of the specific time-dependent protocol or the choice of observables.

\section{Acknowledgement}
We would like to acknowledge the support provided by the project ``Study of quantum chaos and multipartite entanglement using quantum circuits'' sponsored by the Science and Engineering Research Board (SERB), Department of Science and Technology (DST), India under the Core Research Grant CRG/2021/007095. 

\appendix
\section{Appendix: OTOCs using SBOs and Pauli observables with $x$-direction alignment in  Floquet system with periodic field}
\label{periodic_quench_appendix}

Let us discuss the time evolution of the Floquet system, which is defined by the Hamiltonian:  
\begin{eqnarray}
\hat{H}(t) = J\hat{H}_{xx} + h_x \hat{H}_x + h_z \sum_{n=-\infty}^{\infty} \delta\Big(n - \frac{t}{\tau}\Big)\hat{H}_z,
\label{Floquet_hamiltonian}
\end{eqnarray}

In Eq.~(\ref{Floquet_hamiltonian}), the longitudinal and transverse fields, $h_x$ and $h_z$, respectively, are constant.
 \par
The wave function under time evolution is defined as follows:
\begin{equation}
\label{operator}
  \psi(x,t) = \hat{U}(t) \psi(x,0),
\end{equation}
where $\hat{U}(t)$ is the time-evolution operator that evolves the wave function from $t = 0$ to $t = t$. The time-dependent Schrödinger equation is given by
\begin{equation}
\label{SE}
  i\hslash \frac{\partial \psi}{\partial t} = \hat{H} \psi.
\end{equation}

Substituting Eq.~(\ref{operator}) into Eq.~(\ref{SE}), we obtain
\begin{equation}
\label{SE1}
  i\hslash \frac{\partial \hat{U}(t)}{\partial t} = \hat{H} \hat{U}(t).
\end{equation}

Initially, at $t = 0$, the operator satisfies $\hat{U}(0) = 1$. Since the Hamiltonian $\hat{H}$ is Hermitian, the time-evolution operator $\hat{U}(t)$ must be unitary. To prove this, we take the adjoint of Eq.~(\ref{SE1}):
\begin{equation}
\label{SE2}
  -i\hslash \frac{\partial \hat{U}^\dagger(t)}{\partial t} = \hat{U}^\dagger(t) \hat{H}.
\end{equation}

Multiplying Eq.~(\ref{SE1}) on the left by $\hat{U}^\dagger(t)$ and Eq.~(\ref{SE2}) on the right by $\hat{U}(t)$, and then subtracting the two equations, we get:
\begin{equation}
\label{diff}
  \frac{d\big(\hat{U}^\dagger(t) \hat{U}(t)\big)}{dt} = 0.
\end{equation}

From Eq.~(\ref{diff}), it follows that $\hat{U}^\dagger(t) \hat{U}(t)$ is a constant. Using the initial condition $\hat{U}(0) = 1$, this constant must be one:
\begin{equation}
  \hat{U}^\dagger(t) \hat{U}(t) = 1.
\end{equation}

For periodic time $t = n\tau$, Eq.~(\ref{operator}) takes the form for one period:
\begin{equation}
  \psi(x,\tau) = \mathcal{\hat{U}}_x(\tau) \psi(x,0).
\end{equation}

If the unitary operator satisfies the condition $\hat{U}(t + n\tau) = \hat{U}(\tau)$ for $n = 1, 2, \dots$, the generalized form of the equation for $n$ periods is:
\begin{equation}
  \psi(x,n\tau) = [\mathcal{\hat{U}}_x(\tau)]^n \psi(x,0).
\end{equation}

Thus, the time-evolution operator for the periodic system is given by:
\begin{equation}
  \hat{U}(n\tau) = [\mathcal{\hat{U}}_x(\tau)]^n.
\end{equation}
In Floquet systems, fields are applied as kicks in the transverse direction, and the strength of the field is stronger than the interaction field strength $J$. When the kicking transverse field is applied to the system, the interaction strength has no significant effect. Due to this, the Floquet map is expressed as the product of two exponential terms \cite{gritsev2017integrable}. For a single period of time, the Floquet map can be expressed as:
\begin{eqnarray}
\label{Uf}
\mathcal{\hat U(\tau)}= \exp \left(-i(\hat H_{xx} +\hat H_x) \tau\right) \exp(-i\tau h_z \hat \sigma_l^z).
\end{eqnarray}
\begin{figure*}
      \includegraphics[width=.32\linewidth, height=.20\linewidth]{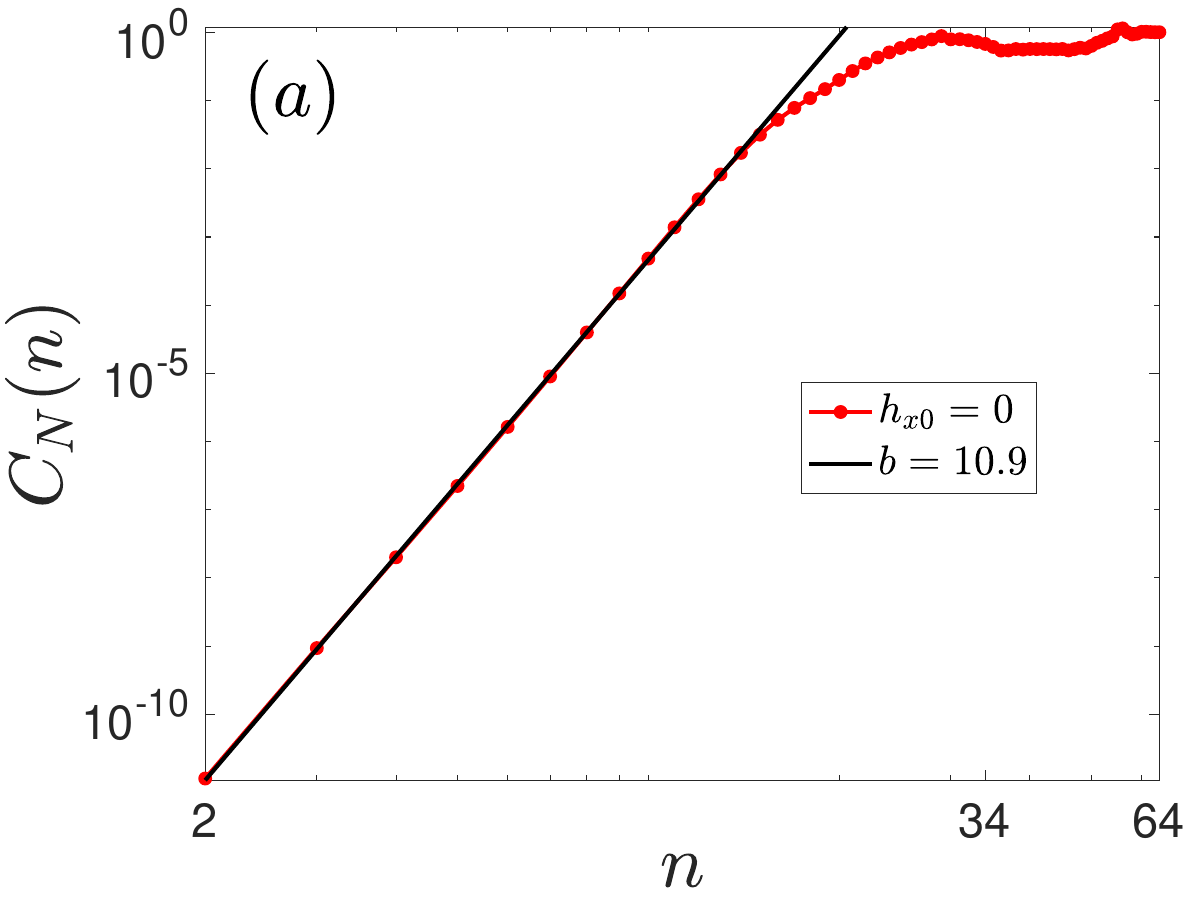}
        \includegraphics[width=.32\linewidth, height=.20\linewidth]{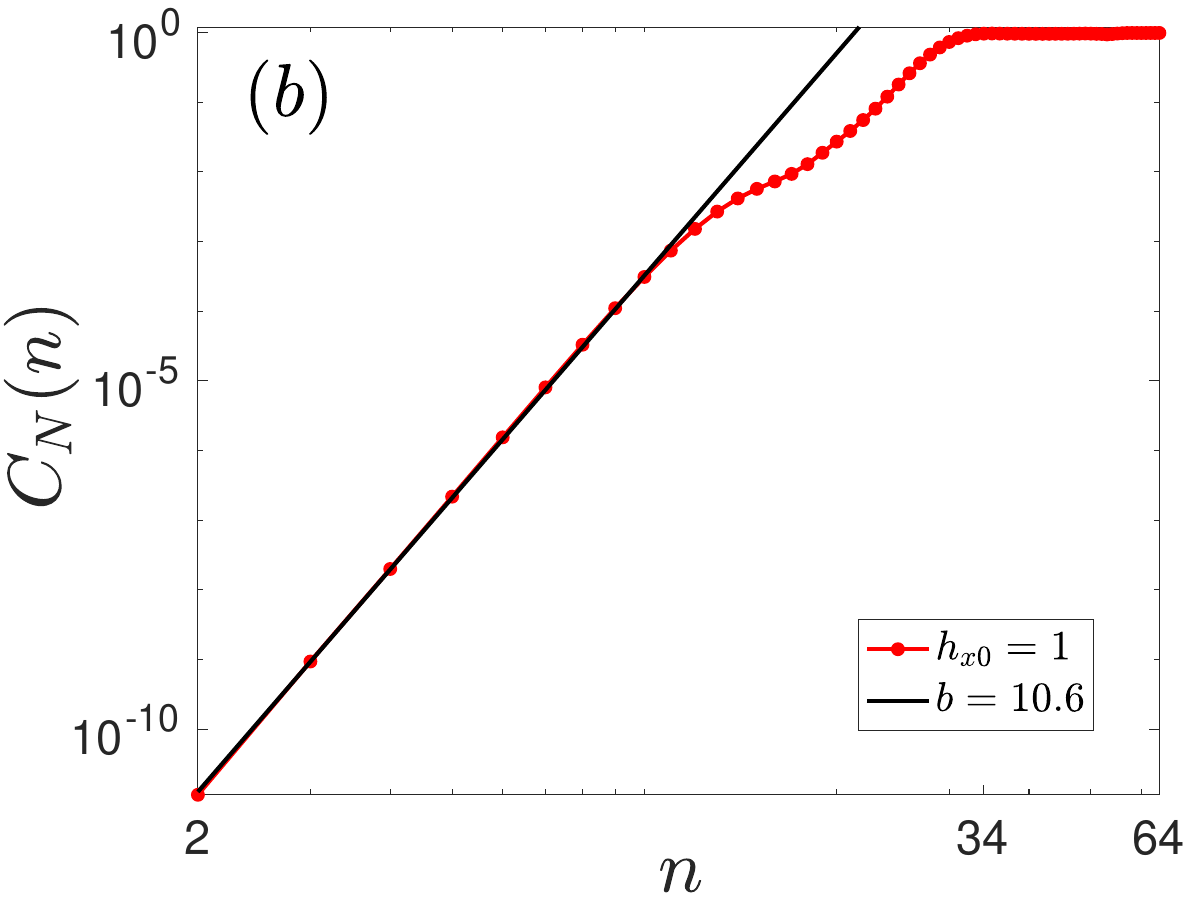}
        \includegraphics[width=.32\linewidth, height=.20\linewidth]{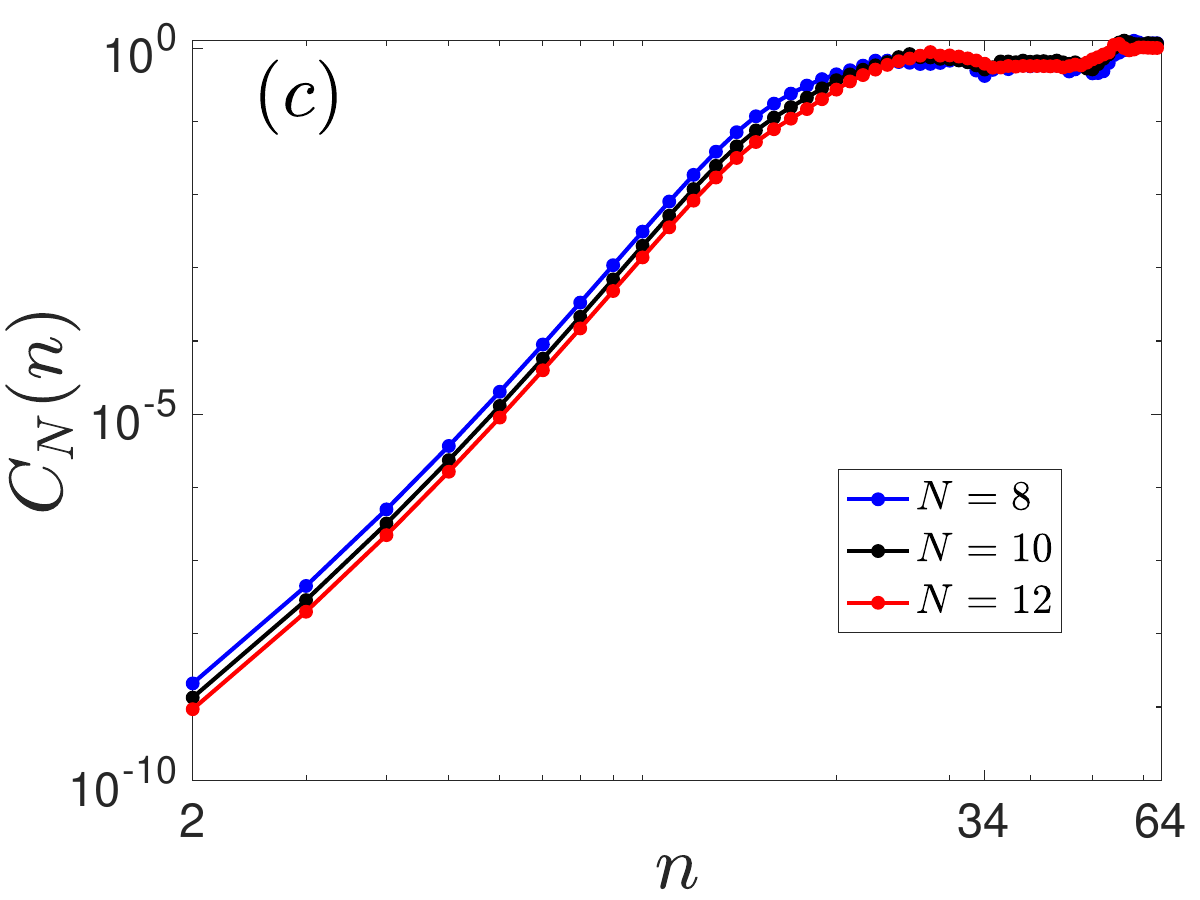}
   \caption{$C(n)/C(\infty)$ using SBO vs. $n$ ($\log-\log$)  for $(a)$ integrable $\mathcal{\hat U}_{0p}$ and $(b)$ nonintegrable $\mathcal{\hat U}_{xp}$ system ($h_{x0}=1$) with system size $N=12$. $(c)$ $C(n)/C(\infty)$ using SBO vs. $n$ ($\log-\log$)  for  integrable $\mathcal{\hat U}_{0p}$  with system size $N=8,10$, and $12$.  Other parameters are: $J=1$, $h_{z}=4$, $t_{\rm max}=8\pi$,  $\tau=\pi/4$, and $\alpha=\pi/2 t_{\rm max}$. Open boundary conditions are considered. } 
     \label{quench_block_hx_rand5_nint_p0_N18}
 \end{figure*}

\begin{figure*}
      \centering
       \includegraphics[width=.32\linewidth, height=.20\linewidth]{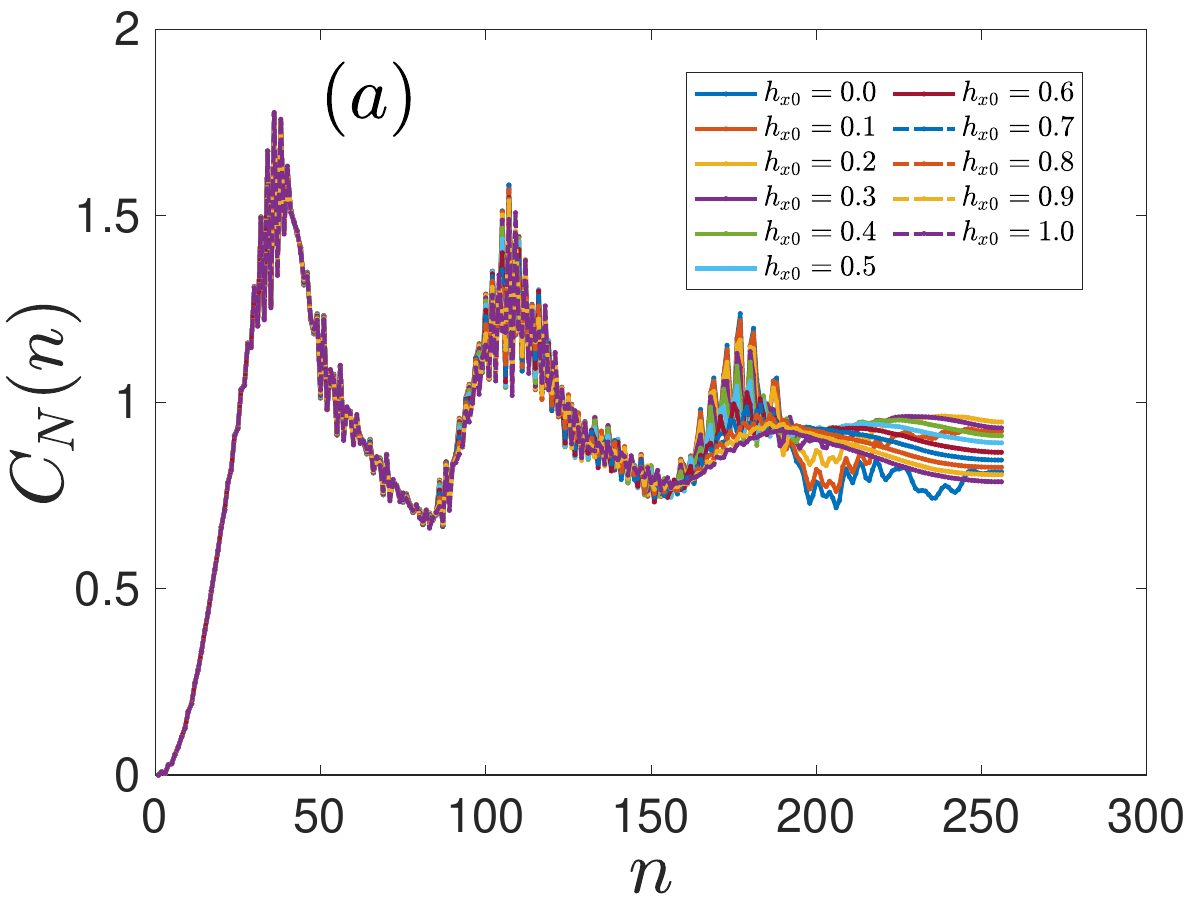}
       \includegraphics[width=.32\linewidth, height=.20\linewidth]{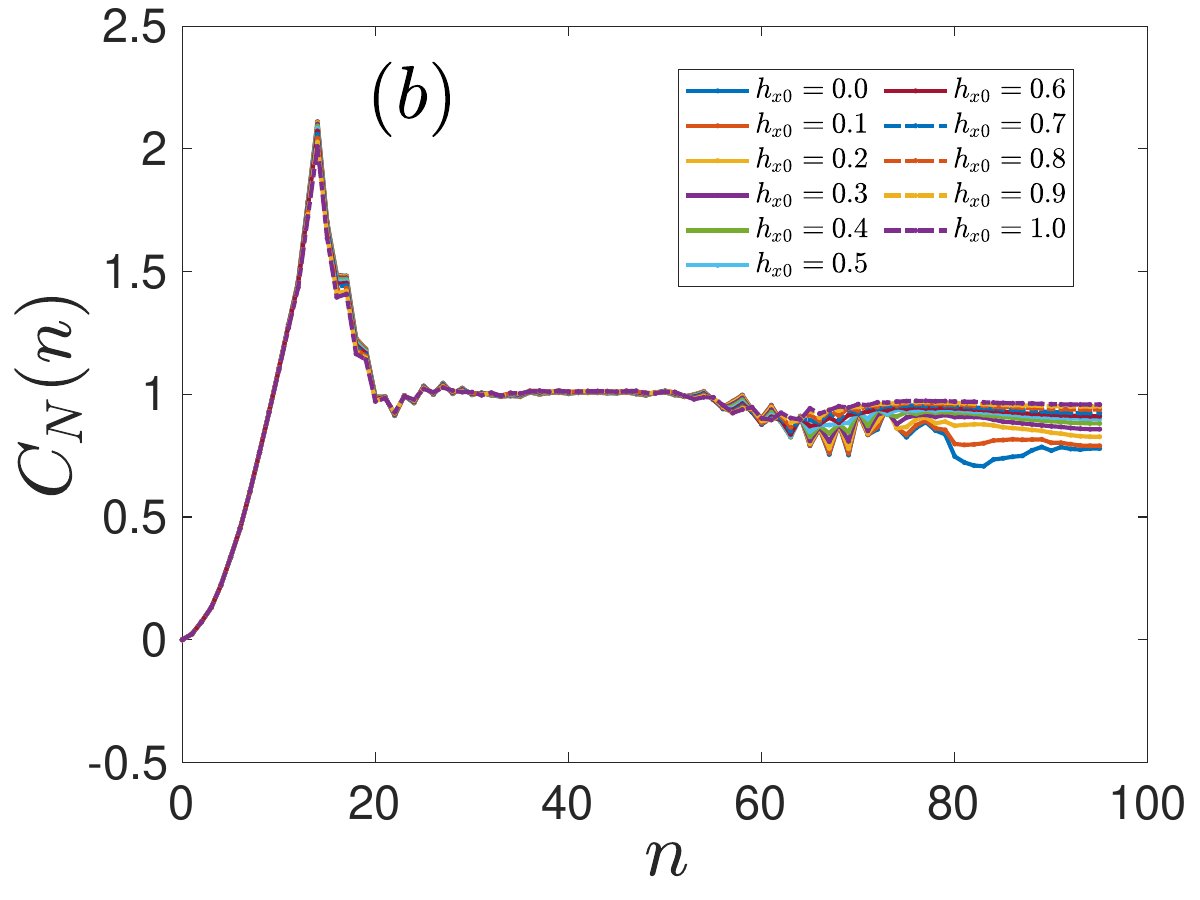}
     \includegraphics[width=.32\linewidth, height=.20\linewidth]{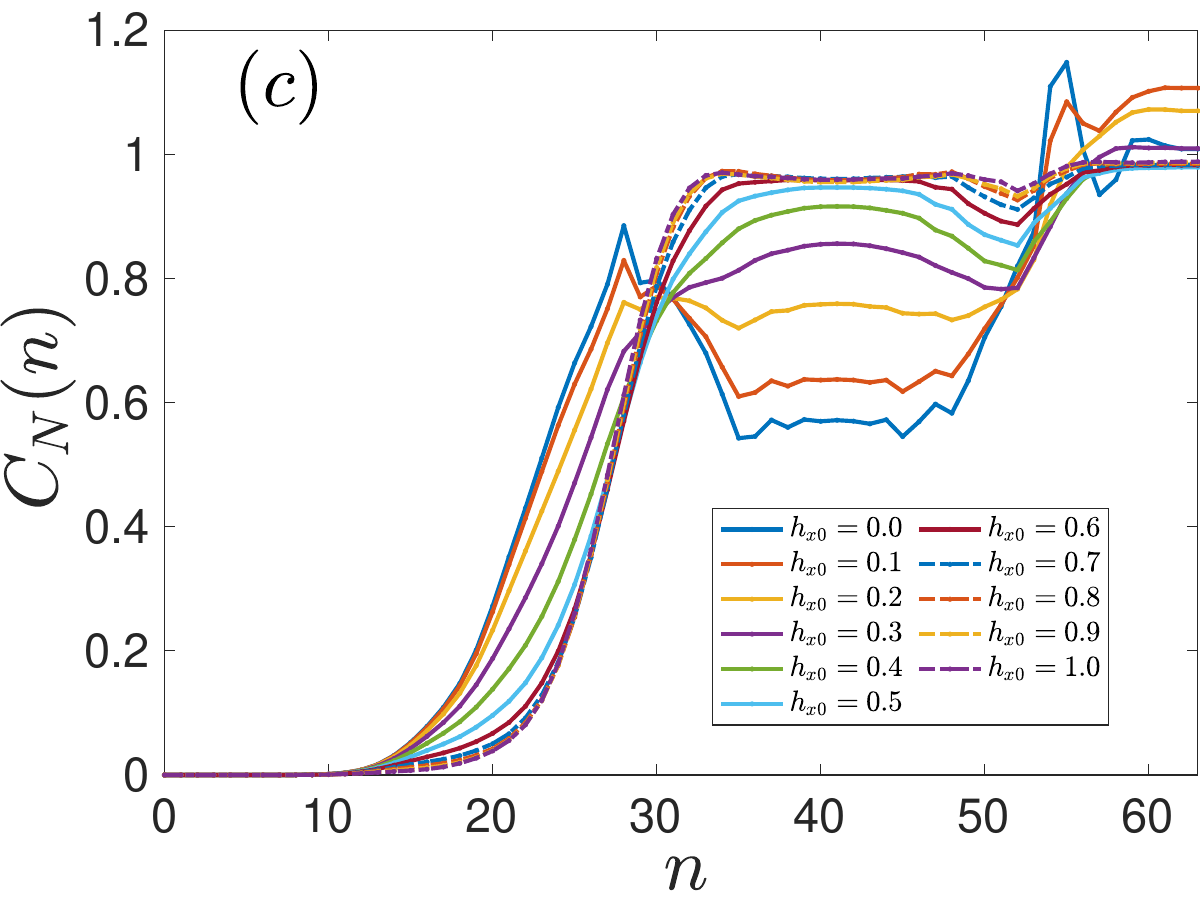}
       \caption{$C_N(n)$ vs $n$ for different value of  $h_{x0}$  lies in between $0$ to $1$ differing by $0.1$ for nonintegrable $\mathcal{\hat U}_{xp}$  system at period $(a)$ $\tau=\pi/16$, $(b)$ $\tau=\pi/6$, and $(c)$ $\tau=\pi/4$. Other parameters are: $J=1$, $h_{z0}=4$, $t_{\rm max}=16\pi$, $\alpha=\pi/2 t_{\rm max}$, and system size $N=12$. Open boundary conditions are considered. }
       \label{OTOC_sat} 
\end{figure*}
 
\begin{figure*}
 \includegraphics[width=.32\linewidth, height=.20\linewidth]{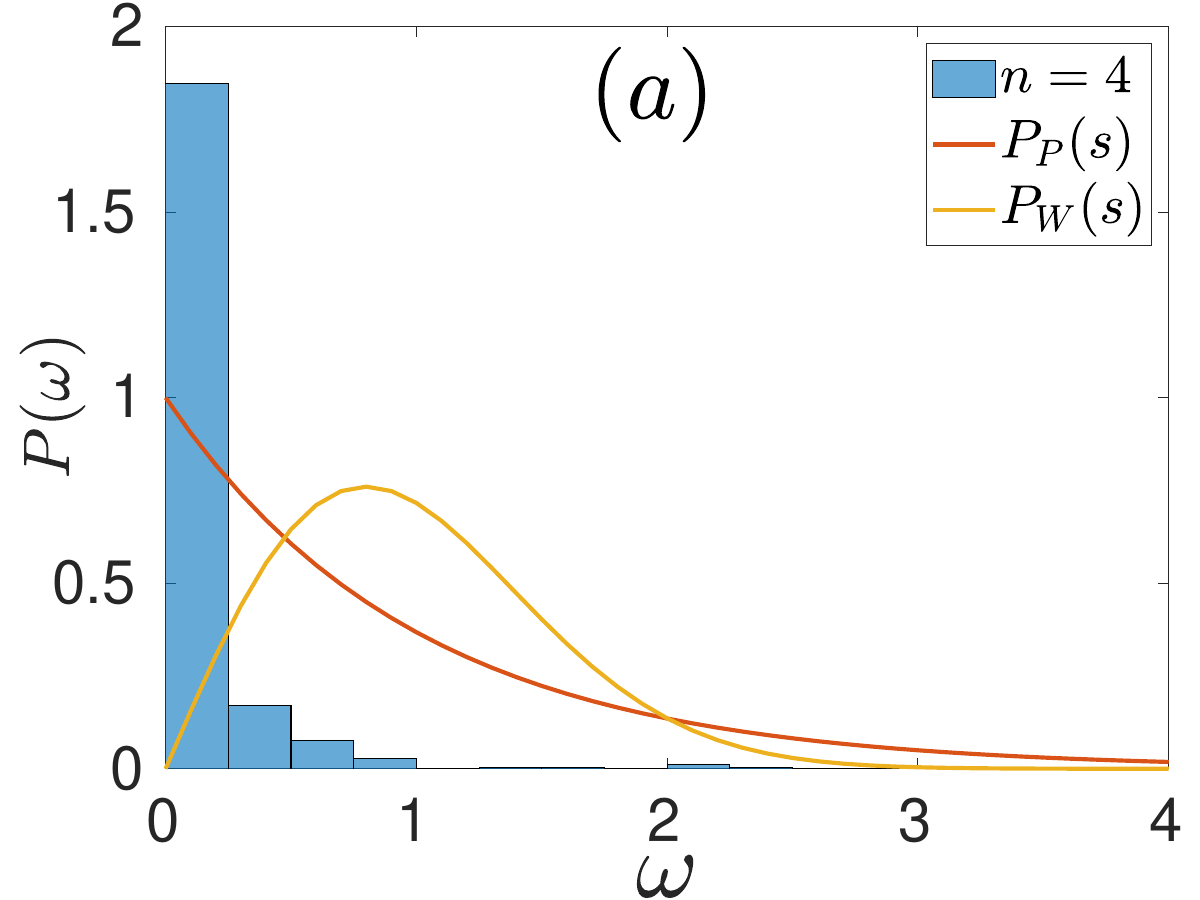}
        \includegraphics[width=.32\linewidth, height=.20\linewidth]{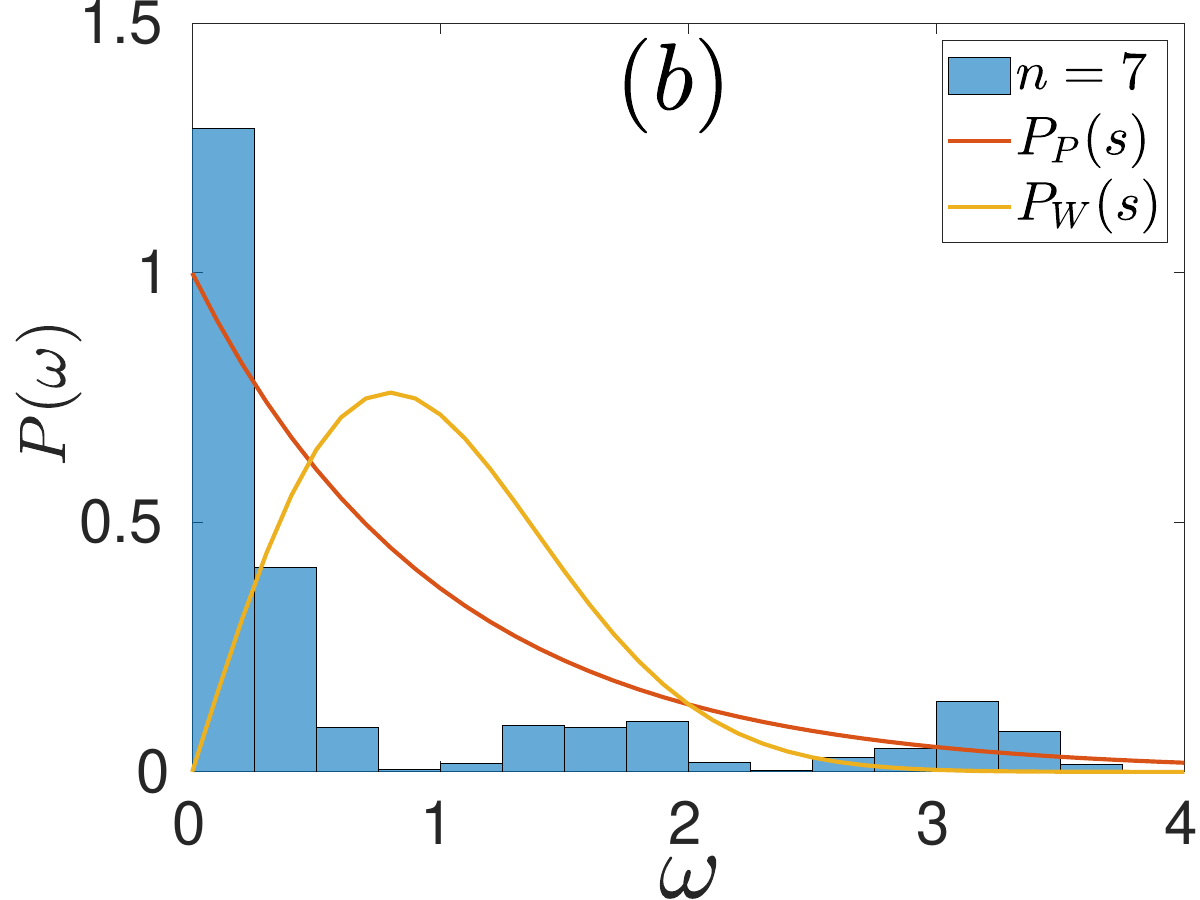}
   \includegraphics[width=.32\linewidth, height=.20\linewidth]{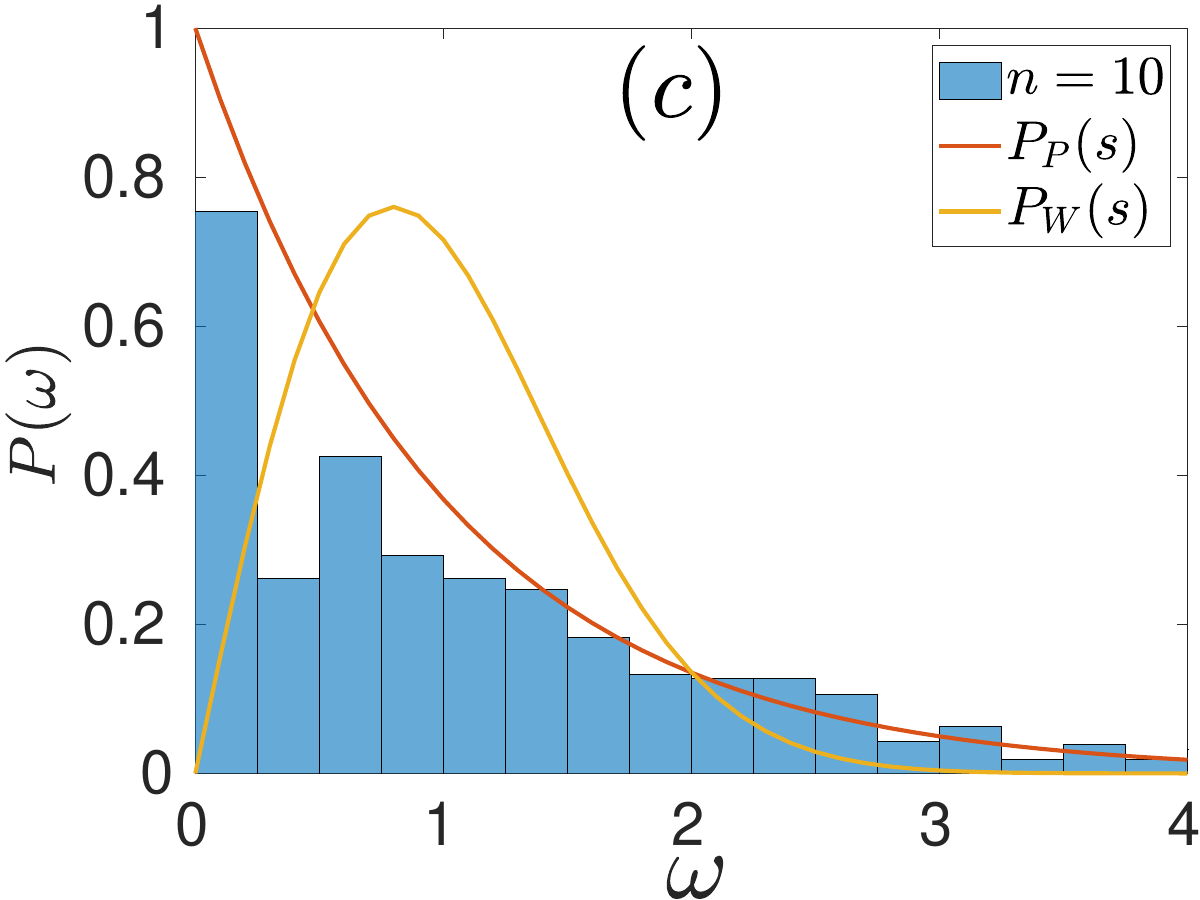}
   \includegraphics[width=.32\linewidth, height=.20\linewidth]{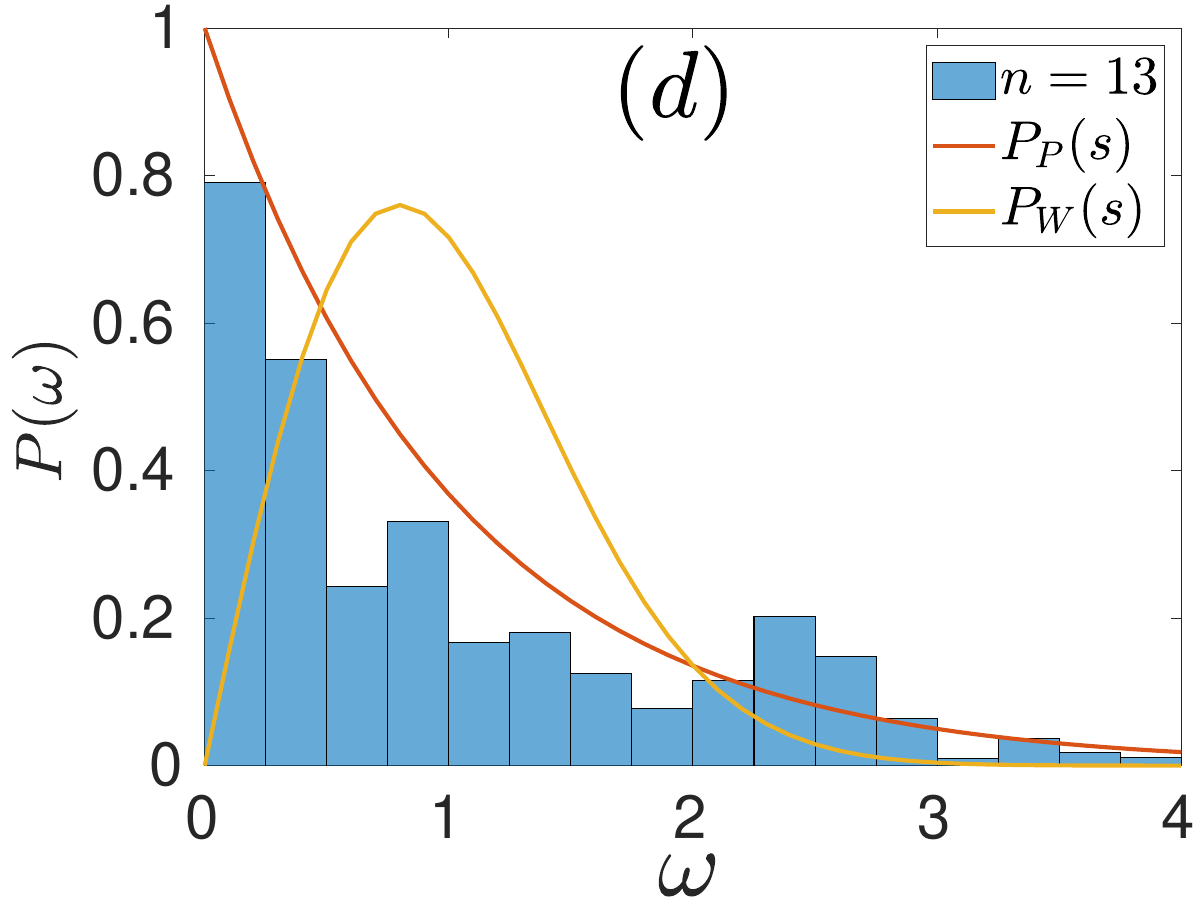}
   \includegraphics[width=.32\linewidth, height=.20\linewidth]{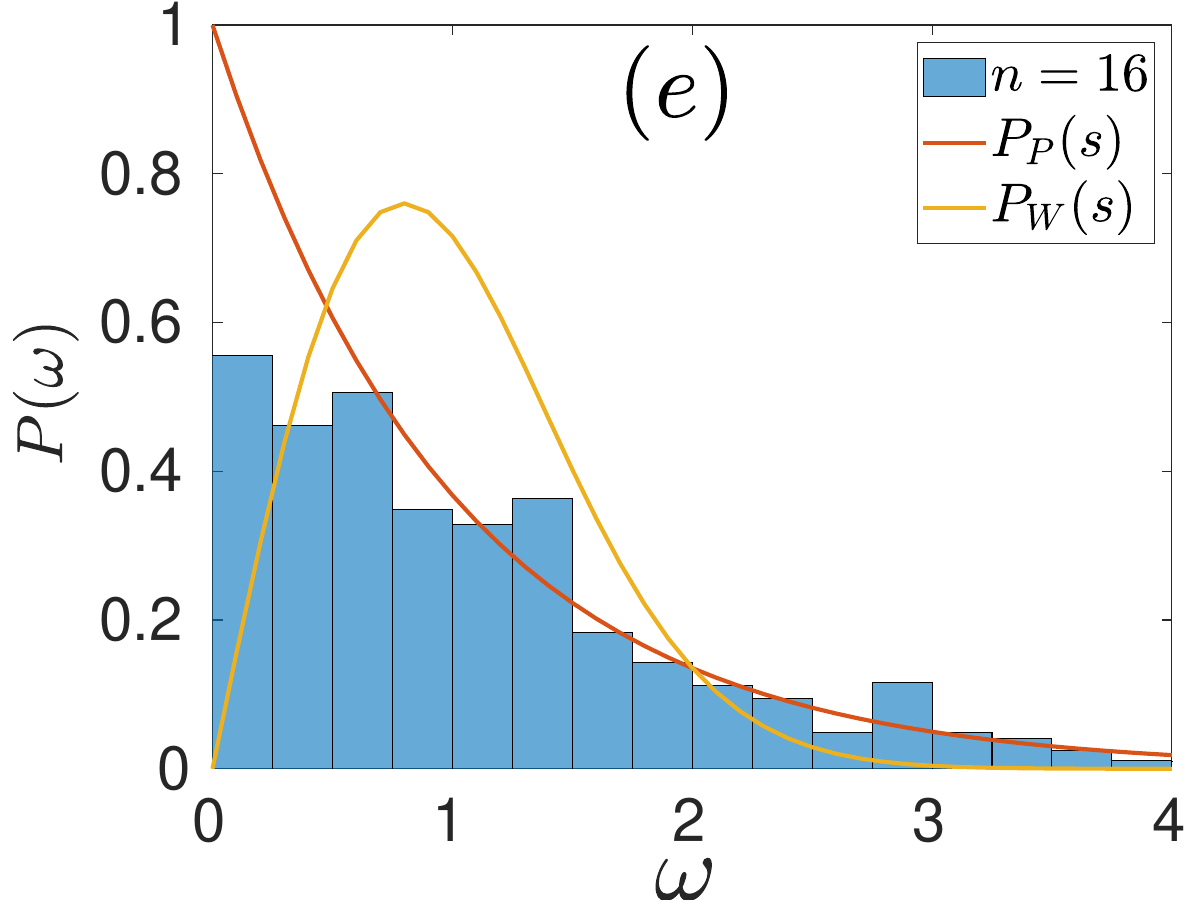}
   \includegraphics[width=.32\linewidth, height=.20\linewidth]{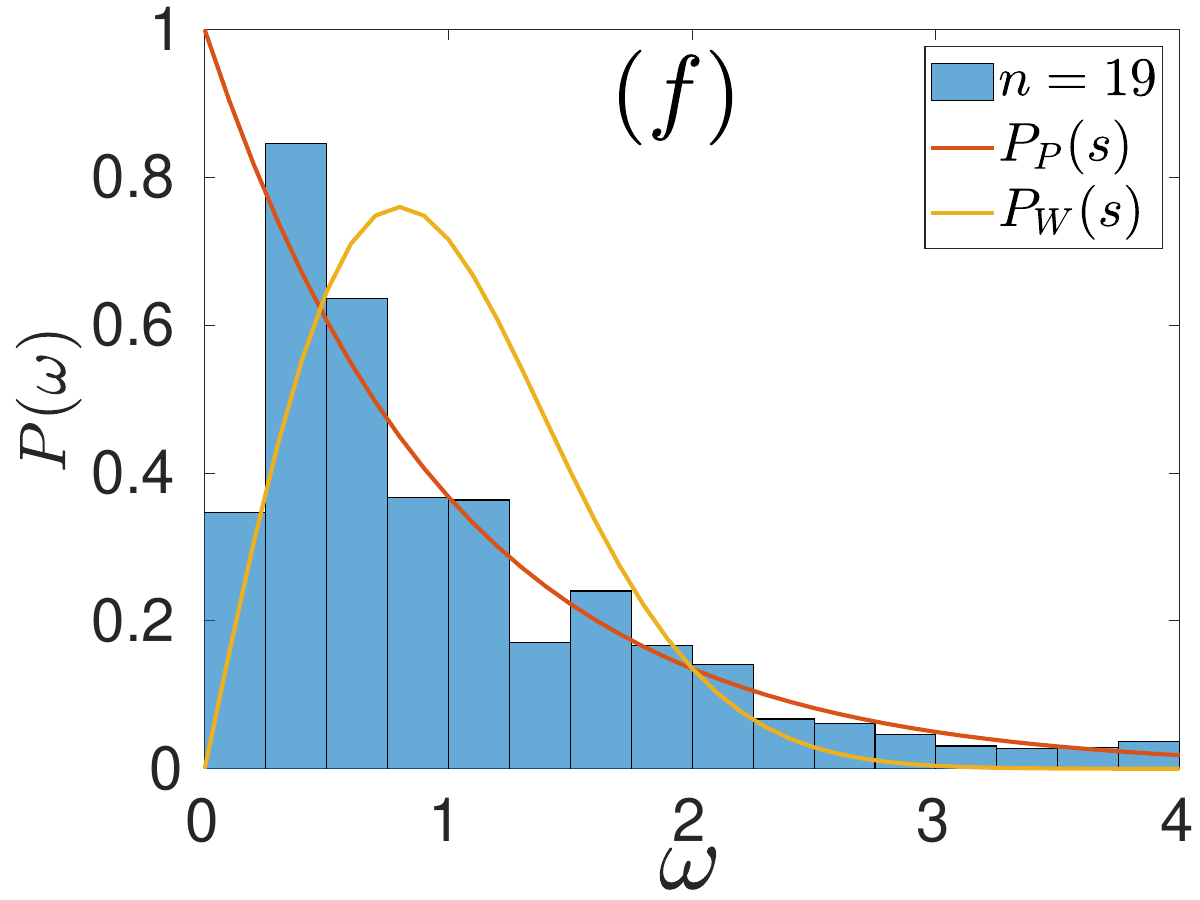}
      \includegraphics[width=.32\linewidth, height=.20\linewidth]{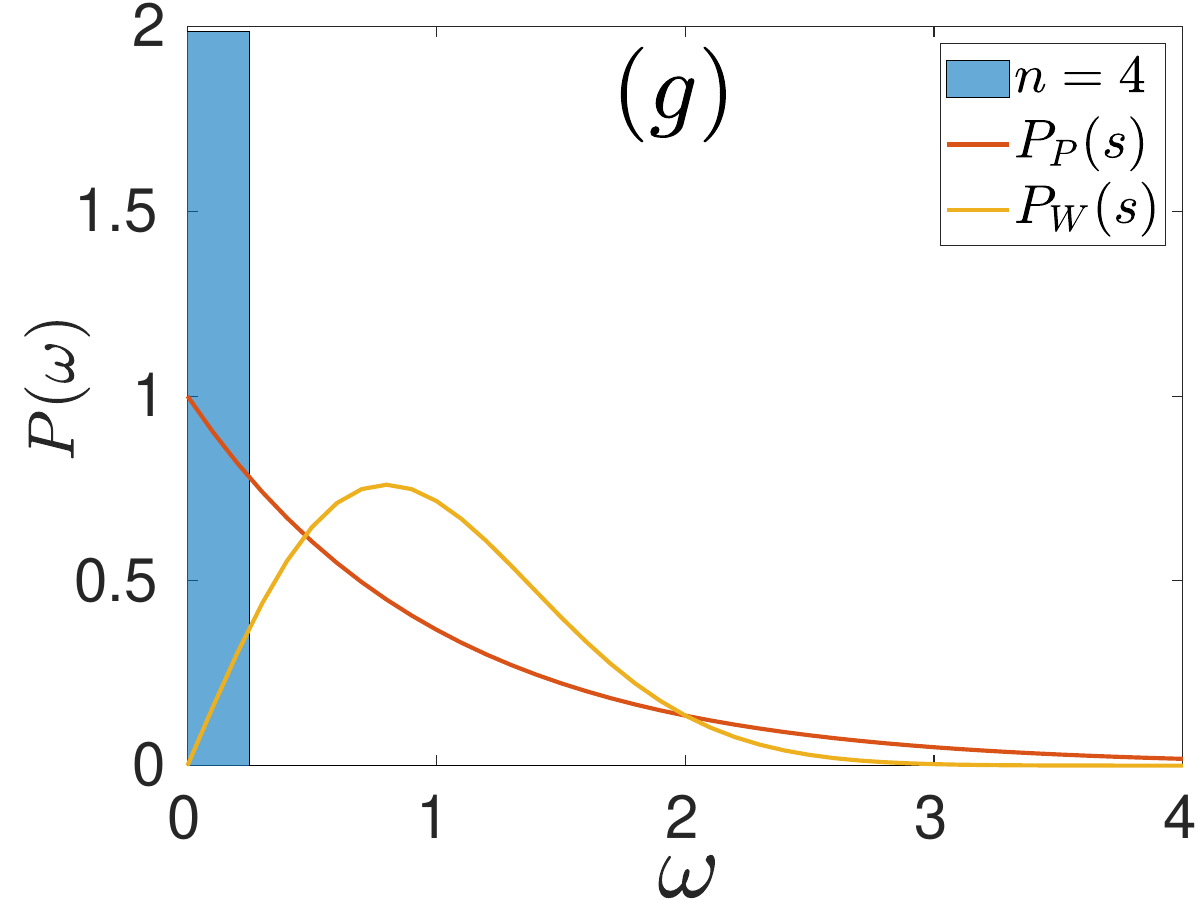}
        \includegraphics[width=.32\linewidth, height=.20\linewidth]{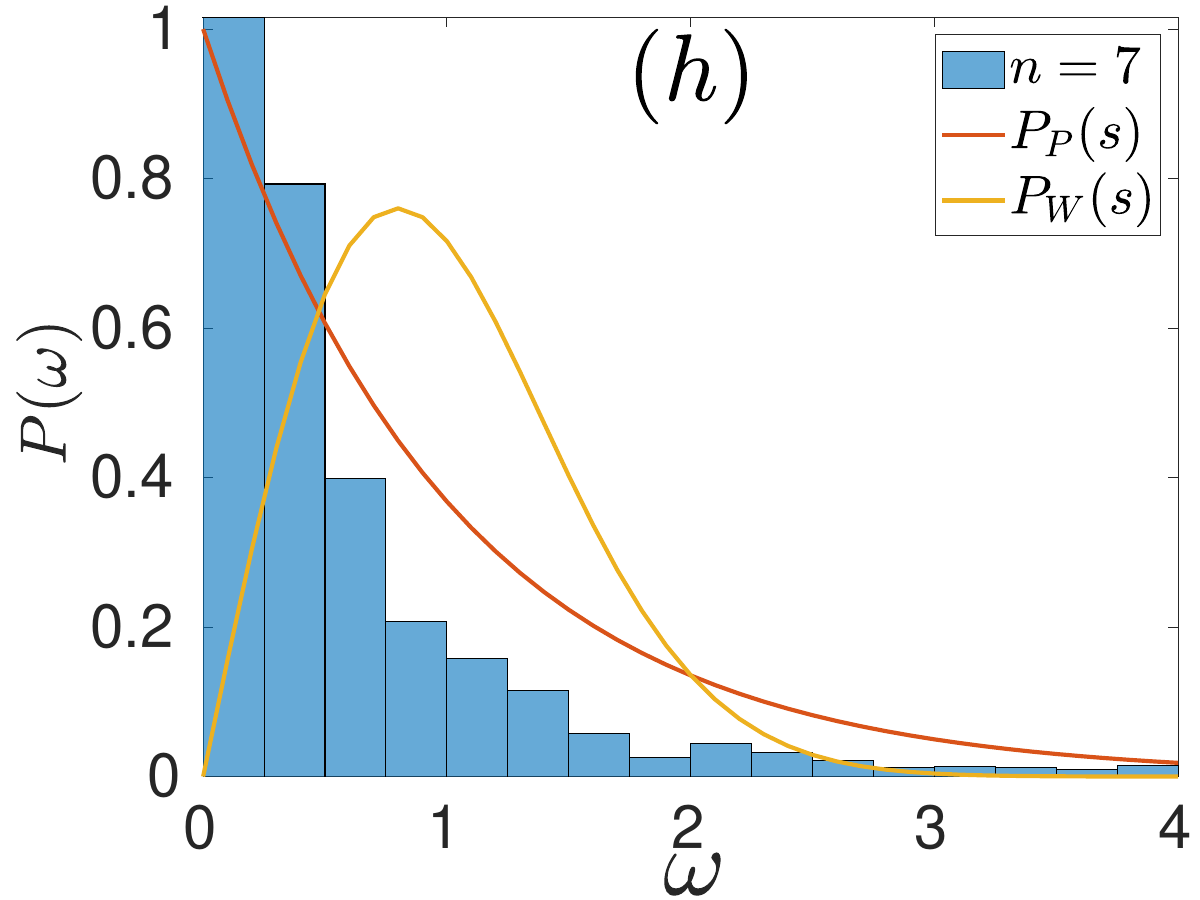}
   \includegraphics[width=.32\linewidth, height=.20\linewidth]{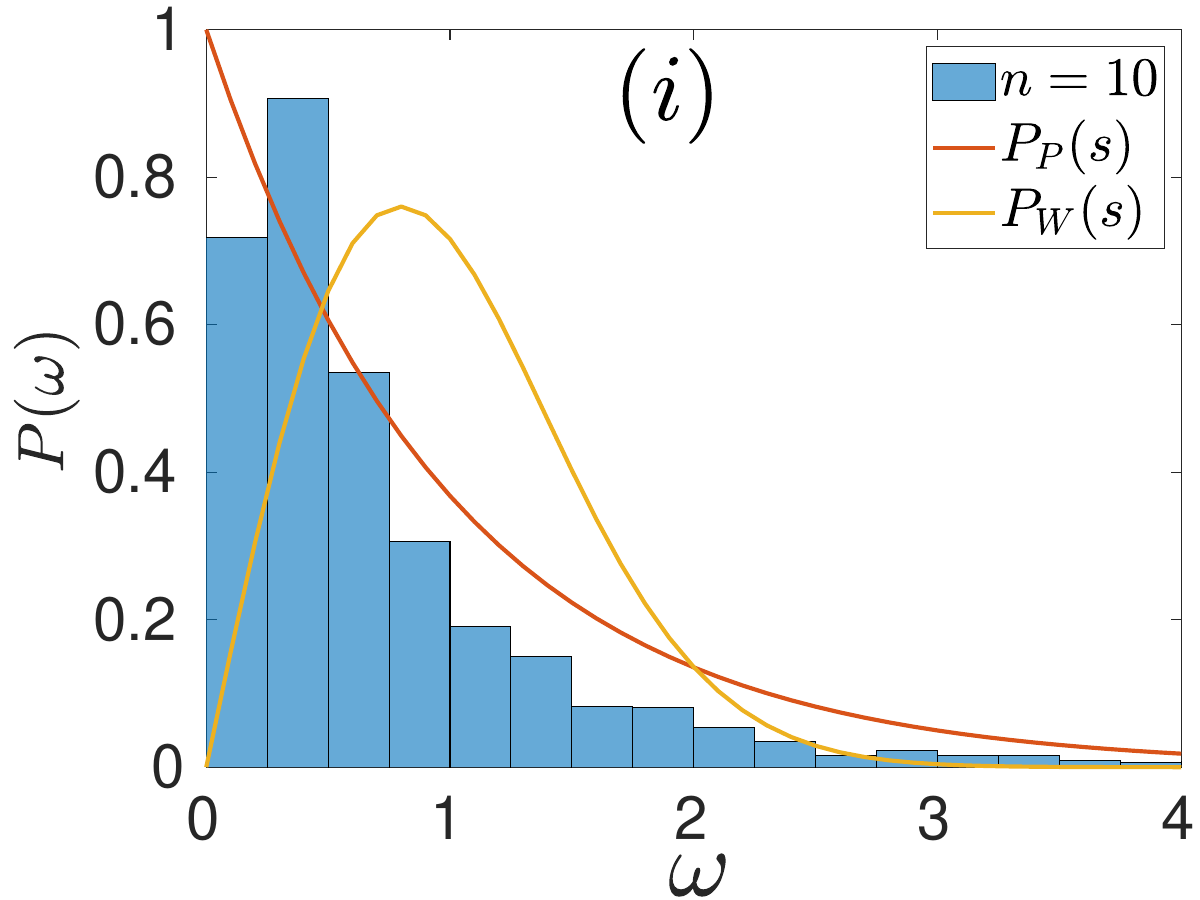}
   \includegraphics[width=.32\linewidth, height=.20\linewidth]{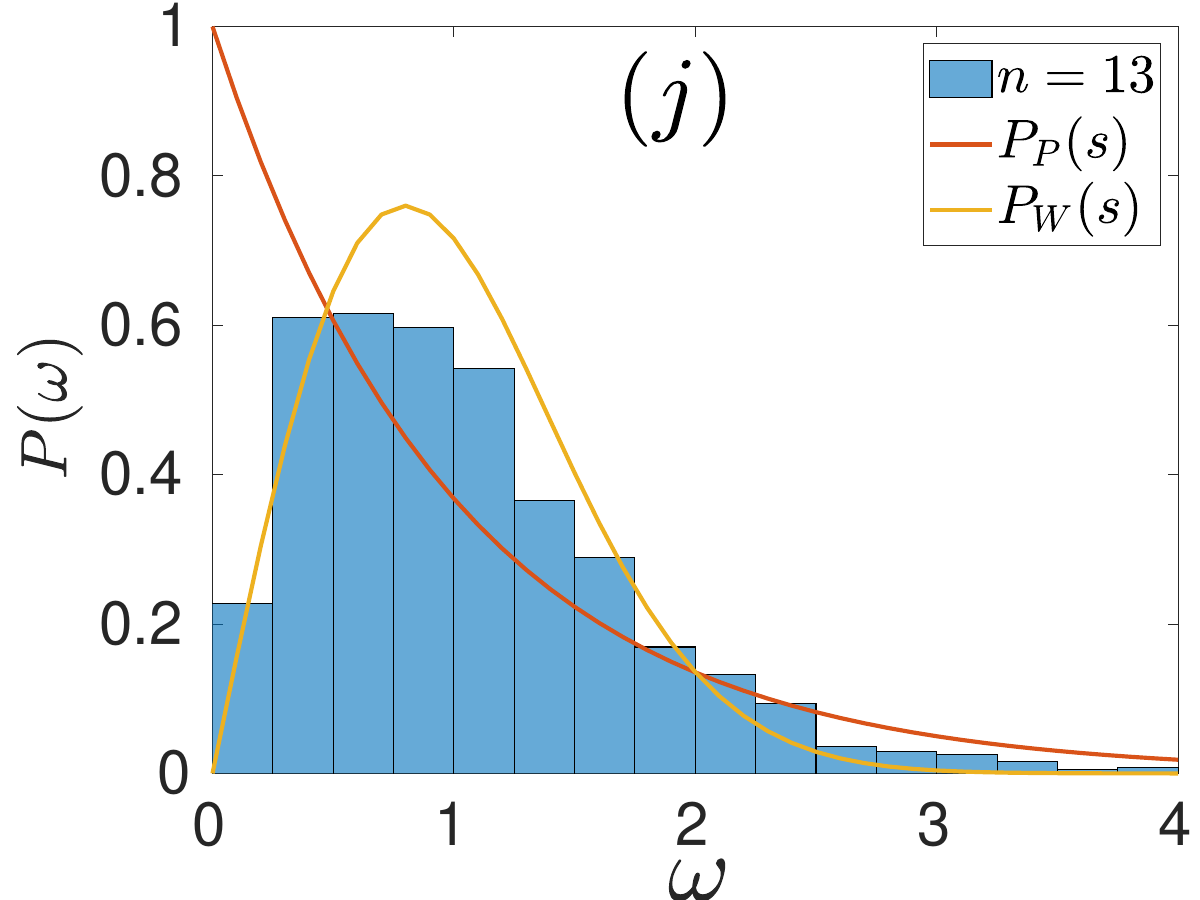}
   \includegraphics[width=.32\linewidth, height=.20\linewidth]{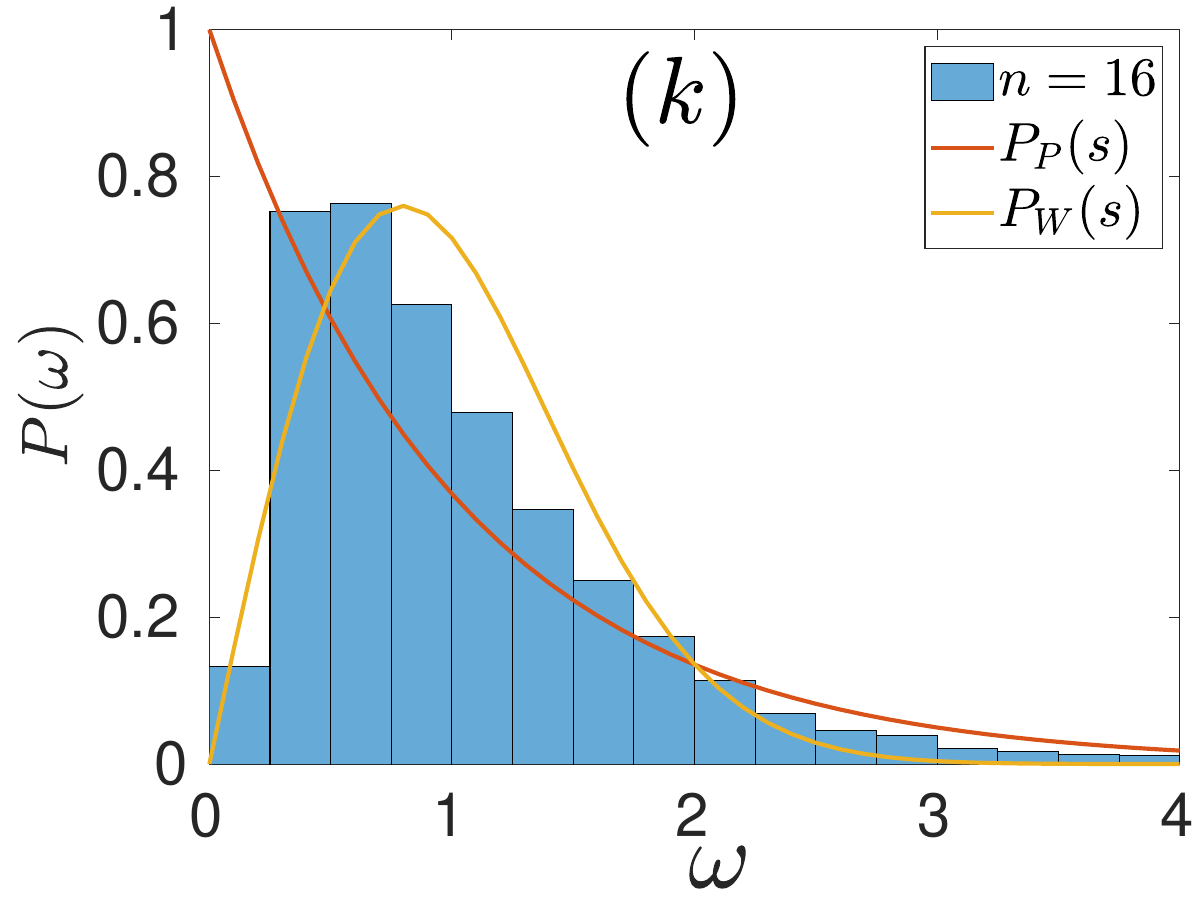}
   \includegraphics[width=.32\linewidth, height=.20\linewidth]{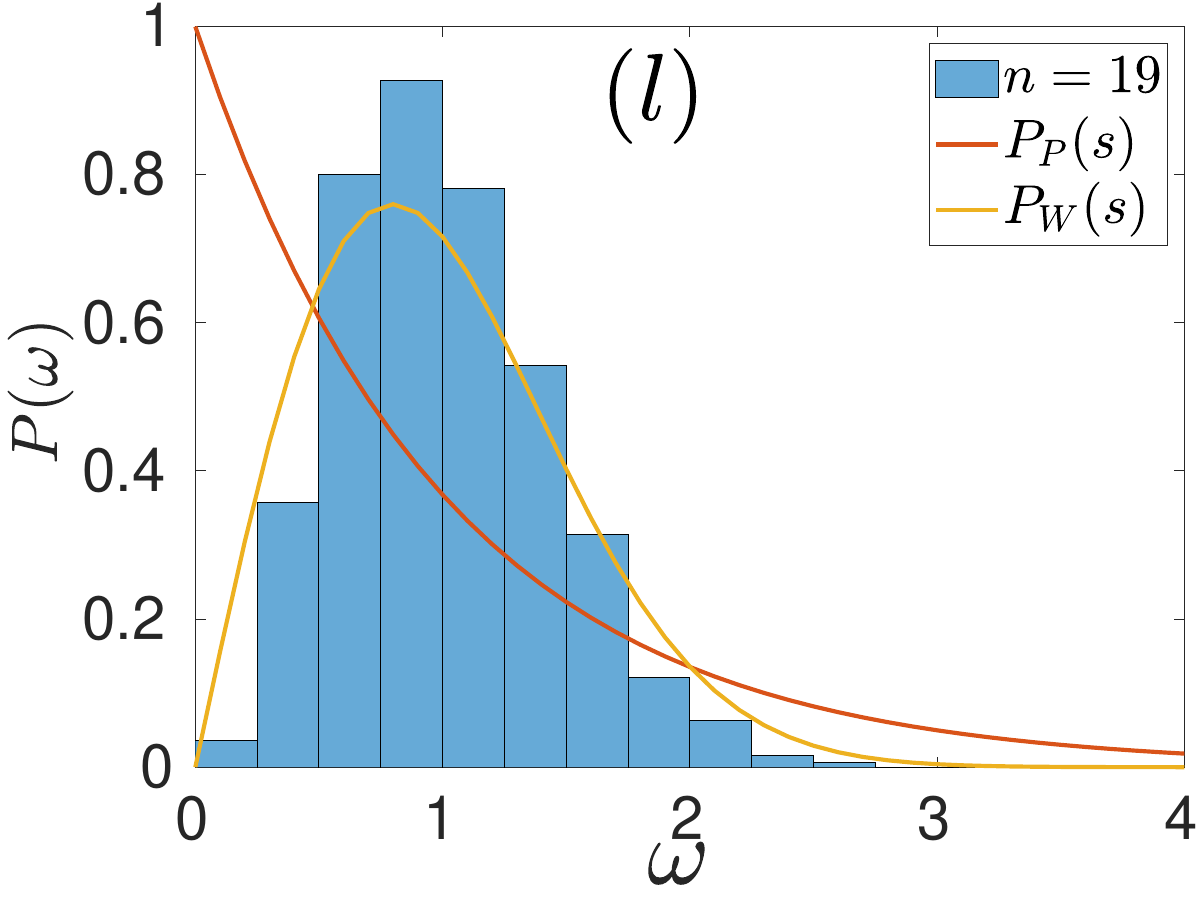}
     \caption{LSS of the ($a-f$) integrable $\mathcal{\hat U}_{0p}$ system, and ($g-l$) nonintegrable $\mathcal{\hat U}_{xp}$ system at period $\tau=\pi/4$ in the even Palindromic space at different kicks mentioned in the legend.  Other parameters are: $\alpha=\pi/2 t_{\rm max}$, and $t_{\rm max}=16\pi$, $J=1$, $h_{x0}=0/4,h_{z0}=4$, $N=12$. Open boundary conditions are considered. }
     \label{periodic_NNSD}
 \end{figure*}

\par
Let us discuss a more challenging scenario where both the longitudinal and transverse magnetic fields are time-dependent. Unlike the previous case, here, the longitudinal field varies with time. We use both longitudinal and transverse magnetic fields, represented by sine and cosine functions, respectively. Mathematically, they are described as $h_x=h_{x0} \sin(\alpha t)$ and $h_z=h_{z0} \cos(\alpha t)$. The corresponding Hamiltonian for these time-dependent fields is given by
\begin{eqnarray}
\hat H(t)=J\hat H_{xx}+h_{x}(t)\hat H_x +h_{z}(t)\sum_{n=-\infty}^{\infty}\delta\Big(n-\frac{t}{\tau}\Big) \hat H_z.
\label{time_hamiltonian}
\end{eqnarray}
\par
The quantum map is the time evolution operator that evolves the system between $n\tau^+$ and $(n+1)\tau^+$ where $\tau^+$ indicates the time just after $\tau$. The quantum map for the time-dependent system is given by $\mathcal{\hat U}_{xp}(n)$ and defined as (for detailed discussion refer to Refs. \cite{mishra2014resonance, Floquet_Notes}) 
\begin{eqnarray}
\label{Uxq_t}
\mathcal{\hat U}_{xp}(n)=&&\left[ \prod_{l=1}^{N}\exp \left(-i \tau\hat \sigma_l^x \hat \sigma_{l+1}^x - \int_{n \tau}^{(n+1)\tau}h_x(t) dt \hat \sigma_l^x \right) \right] \nonumber \\ &&\left[\prod_{l=1}^{N}\exp(-i\tau h_z(n\tau) \hat \sigma_l^z)\right].
\end{eqnarray}
 In the  Floquet system, fields are constant, and no integration is required, unlike in systems with other time-dependent fields \cite{shukla2021, shukla2022characteristic, shukla2022out, naik2019controlled}. However, in this case, we introduce a time-dependent longitudinal field. To calculate the unitary operator under such conditions, we must consider small time intervals (bins) between the transverse kicks. To account for the time dependence of the longitudinal field, it is necessary to integrate the longitudinal field over the interval between two kicks. Hereafter, we use the notation $\mathcal{\hat U}_{xp}$ and $\mathcal{\hat U}_{0p}$ for the nonintegrable and integrable cases, respectively. For the calculation of OTOC, we consider the nonlocalized observables $\hat W_B$ and $\hat V_B$.  At the initial time, $t=0$, the operators $\hat W_B$ and $\hat V_B$ commute. However, immediately after $n$ kicks at $t=n\alpha\tau$, the operator $\hat W_B(n)=\Big(\prod_{k=1}^{n} \hat U(k) \Big)^{\dagger}\hat W_B \Big(\prod_{k=1}^{n}\hat U(k)\Big)$ loses its commutative property. As time progresses, the envelope of the kicked transverse field gradually decreases from its maximum value $h_{z0}$ due to the cosine function. Concurrently, during the same period, the longitudinal field undergoes an increase from $0$ to its maximum value $h_{x0}$ due to the sine functions. The frequency of these fields is held constant at $\alpha=\pi/2t_{\rm max}$, with $t_{\rm max}$ denoting the duration of the evolution, and in this context, $t_{\rm max}$ is taken as $16\pi$.

\begin{figure*}
      \centering
       \includegraphics[width=.32\linewidth, height=.20\linewidth]{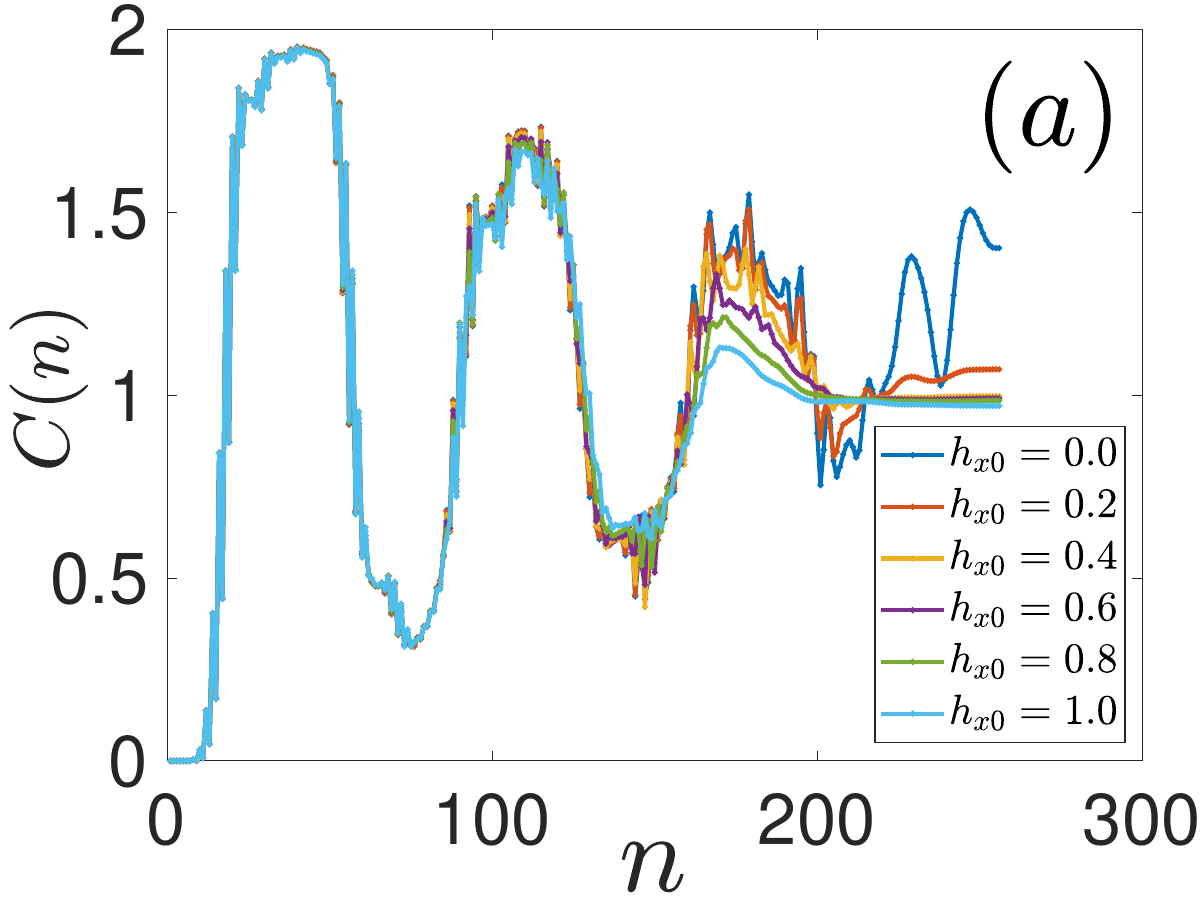}
       \includegraphics[width=.32\linewidth, height=.20\linewidth]{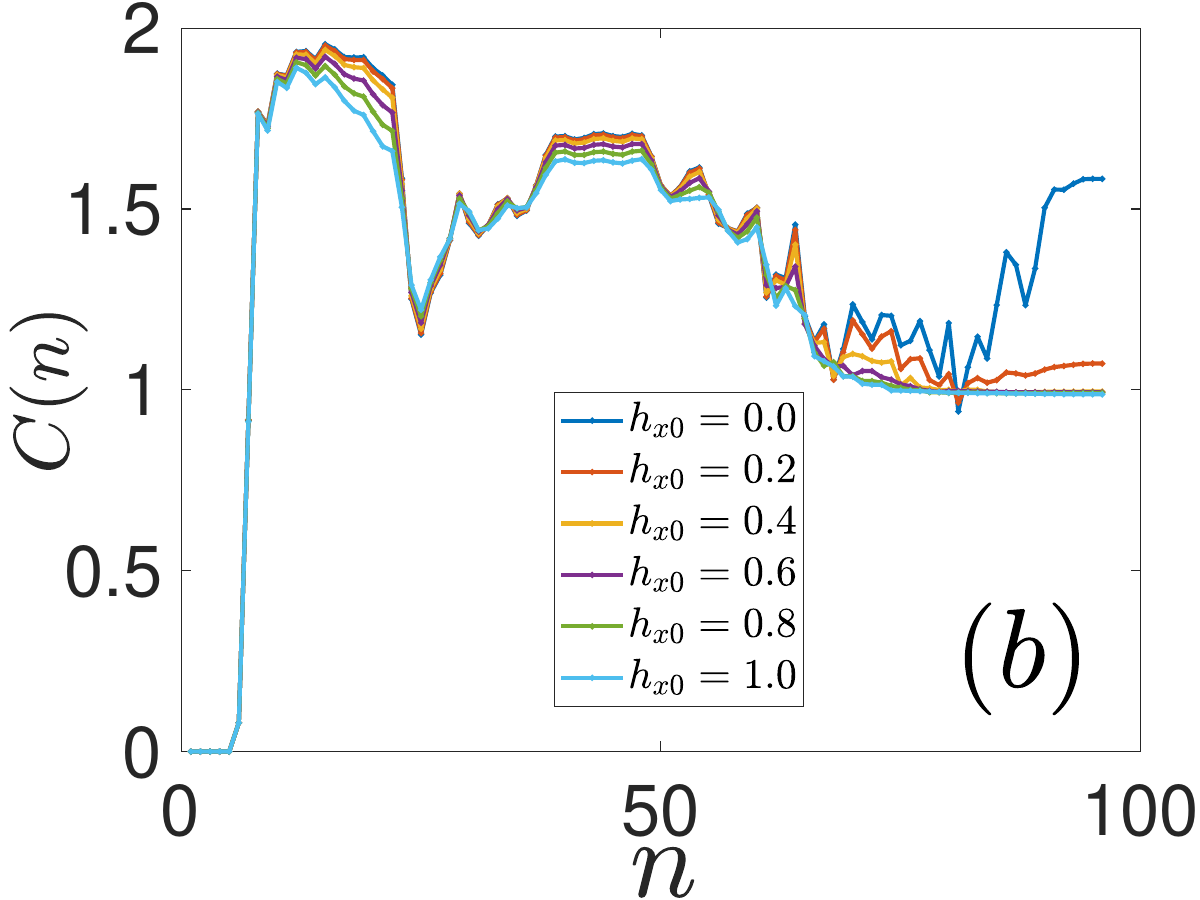}
     \includegraphics[width=.32\linewidth, height=.20\linewidth]{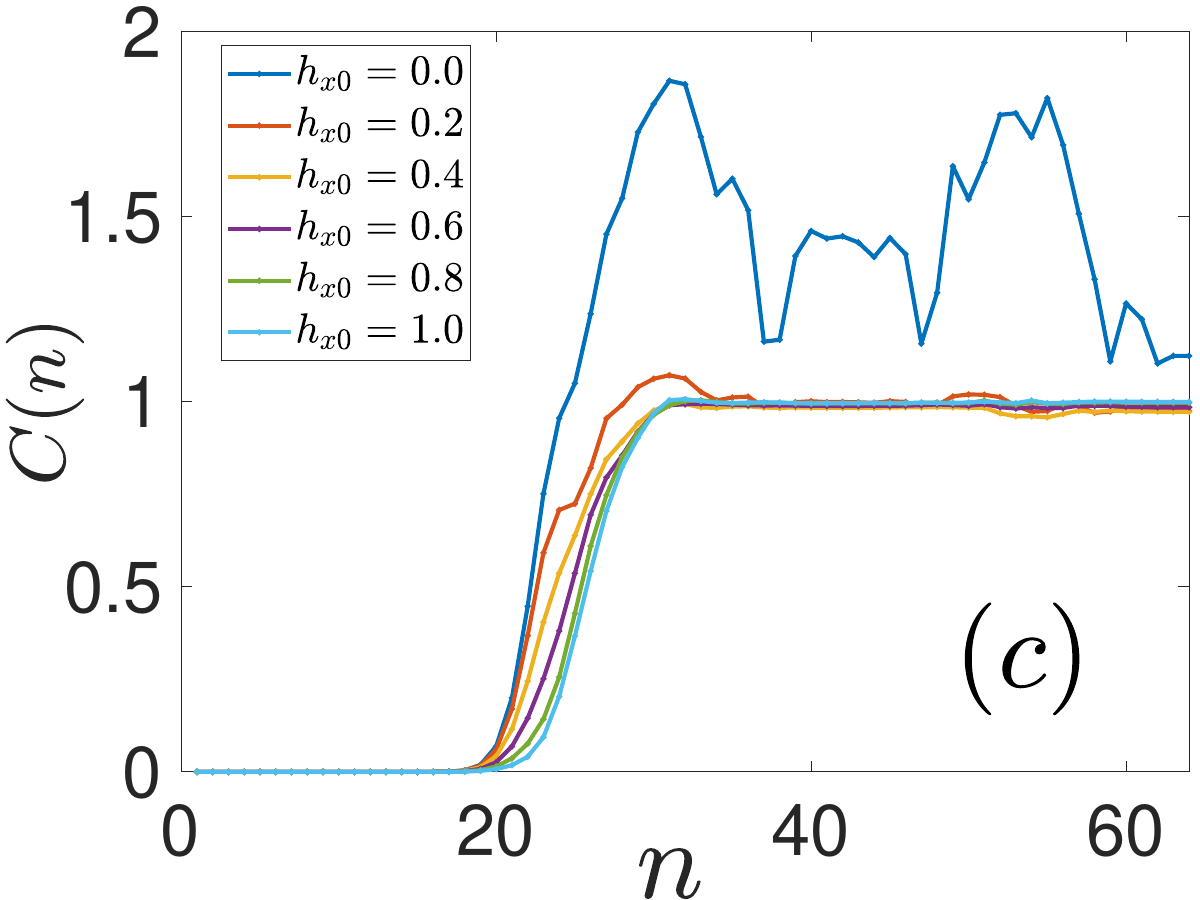}
       \caption{$C(n)$ ($\hat W=\hat S_1^x$, and $\hat V=\hat S_{N/2}^x$) vs $n$ for different value of  $h_{x0}$  lies in between $0$ to $1$ differing by $0.1$ for nonintegrable $\mathcal{\hat U}_{xp}$  system at period $(a)$ $\tau=\pi/16$, $(b)$ $\tau=\pi/6$, and $(c)$ $\tau=\pi/4$. Other parameters are: $J=1$, $h_{z0}=4$, $t_{\rm max}=16\pi$, $\alpha=\pi/2 t_{\rm max}$, and system size $N=12$. Open boundary conditions are considered. }
       \label{sx_OTOC_sat} 
\end{figure*}

\begin{figure*}
      \centering
       \includegraphics[width=.32\linewidth, height=.20\linewidth]{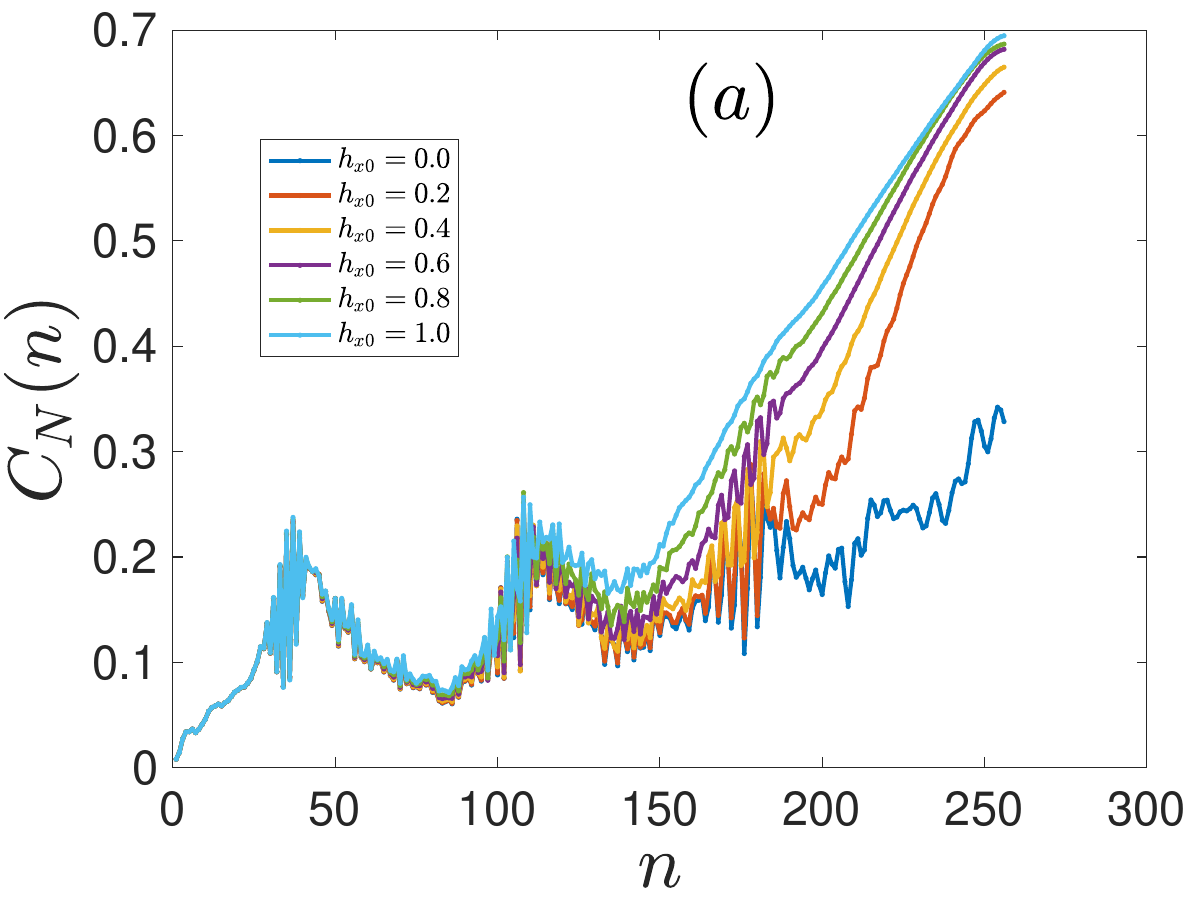}
       \includegraphics[width=.32\linewidth, height=.20\linewidth]{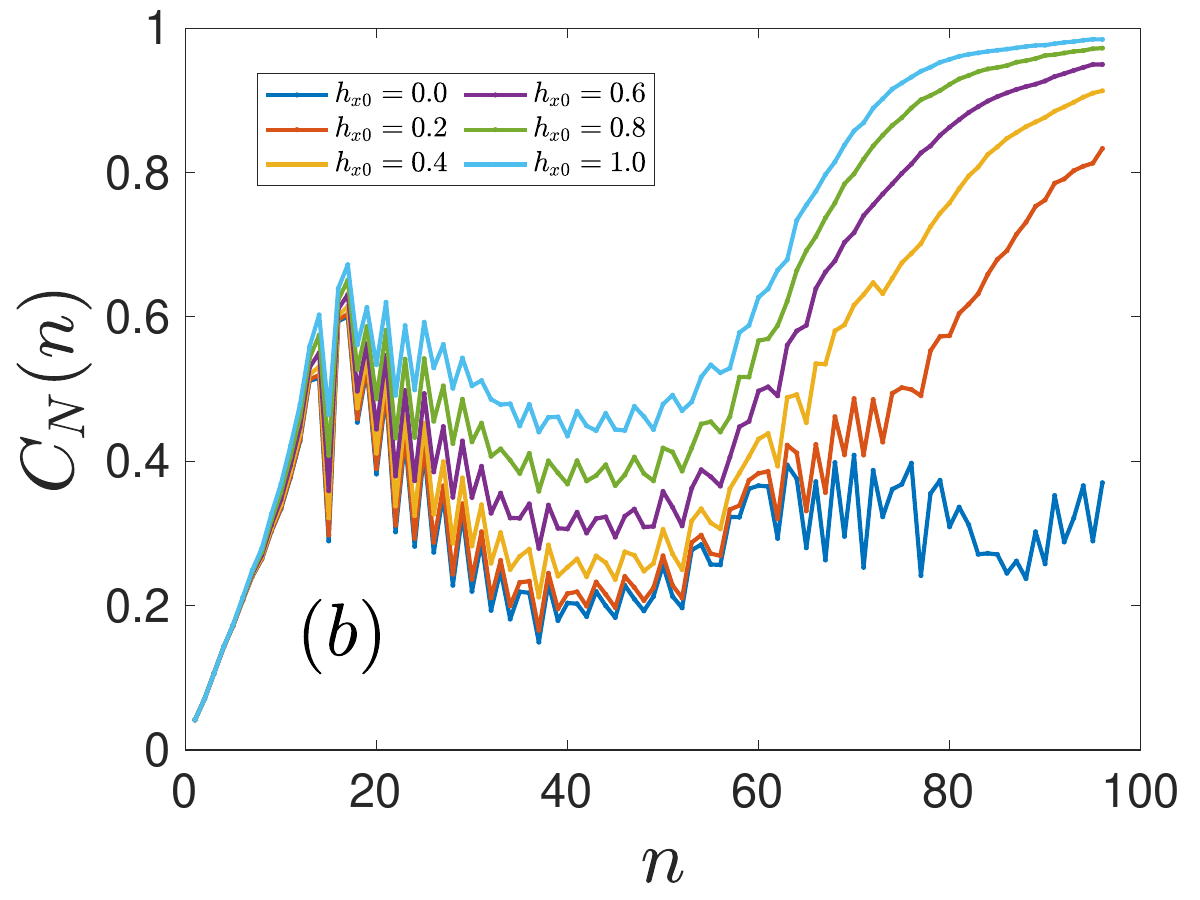}
     \includegraphics[width=.32\linewidth, height=.20\linewidth]{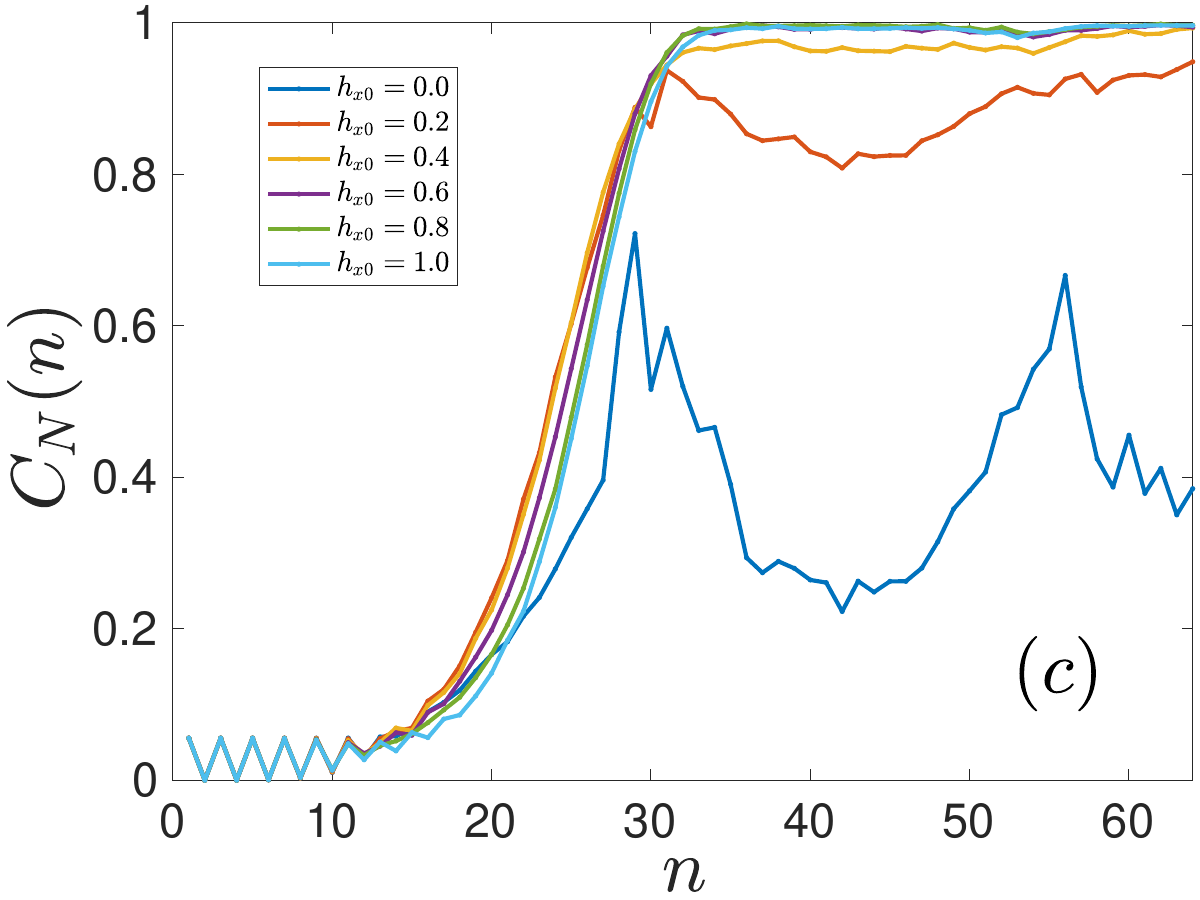}
       \caption{$C_N(n)$ ($\hat W=\sum_{i=1}^{N/2}\hat S_i^z$, and $\hat V=\sum_{i=N/2+1}^{N}\hat S_i^z$) vs $n$ for different value of  $h_{x0}$  lies in between $0$ to $1$ differing by $0.2$ for nonintegrable $\mathcal{\hat U}_{xp}$  system at period $(a)$ $\tau=\pi/16$, $(b)$ $\tau=\pi/6$, and $(c)$ $\tau=\pi/4$. Other parameters are: $J=1$, $h_{z0}=4$, $t_{\rm max}=16\pi$, $\alpha=\pi/2 t_{\rm max}$, and system size $N=12$. Open boundary conditions are considered. }
       \label{Block_z_OTOC_sat} 
\end{figure*}

\begin{figure*}
      \centering
       \includegraphics[width=.32\linewidth, height=.20\linewidth]{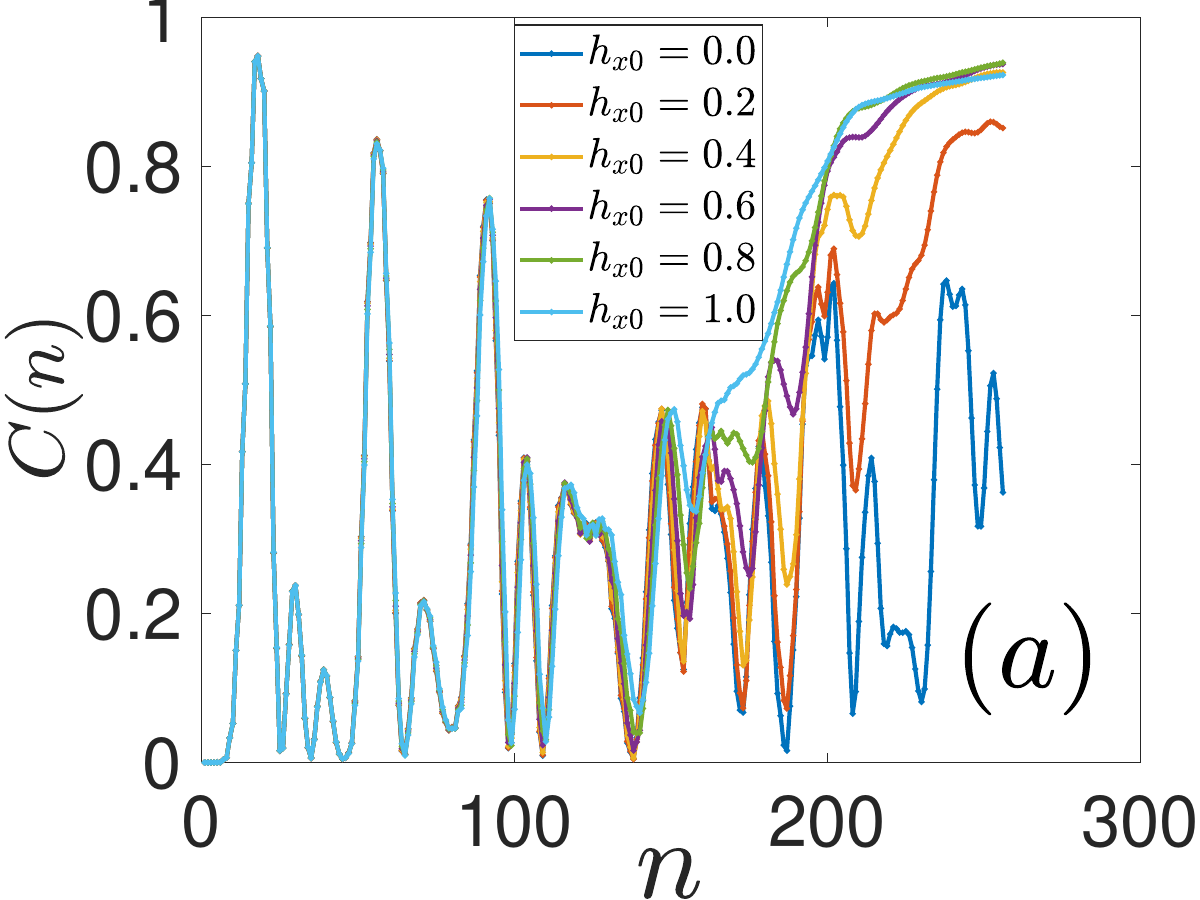}
       \includegraphics[width=.32\linewidth, height=.20\linewidth]{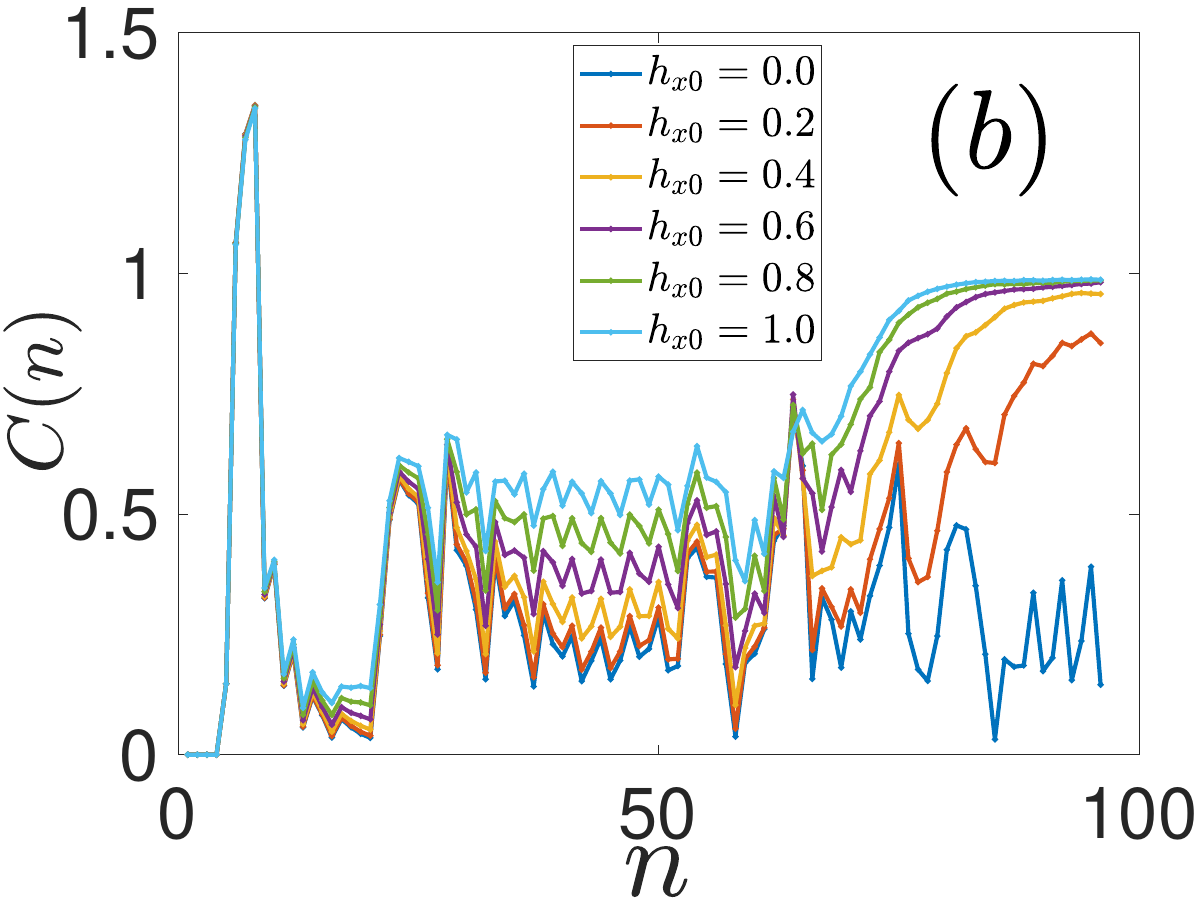}
     \includegraphics[width=.32\linewidth, height=.20\linewidth]{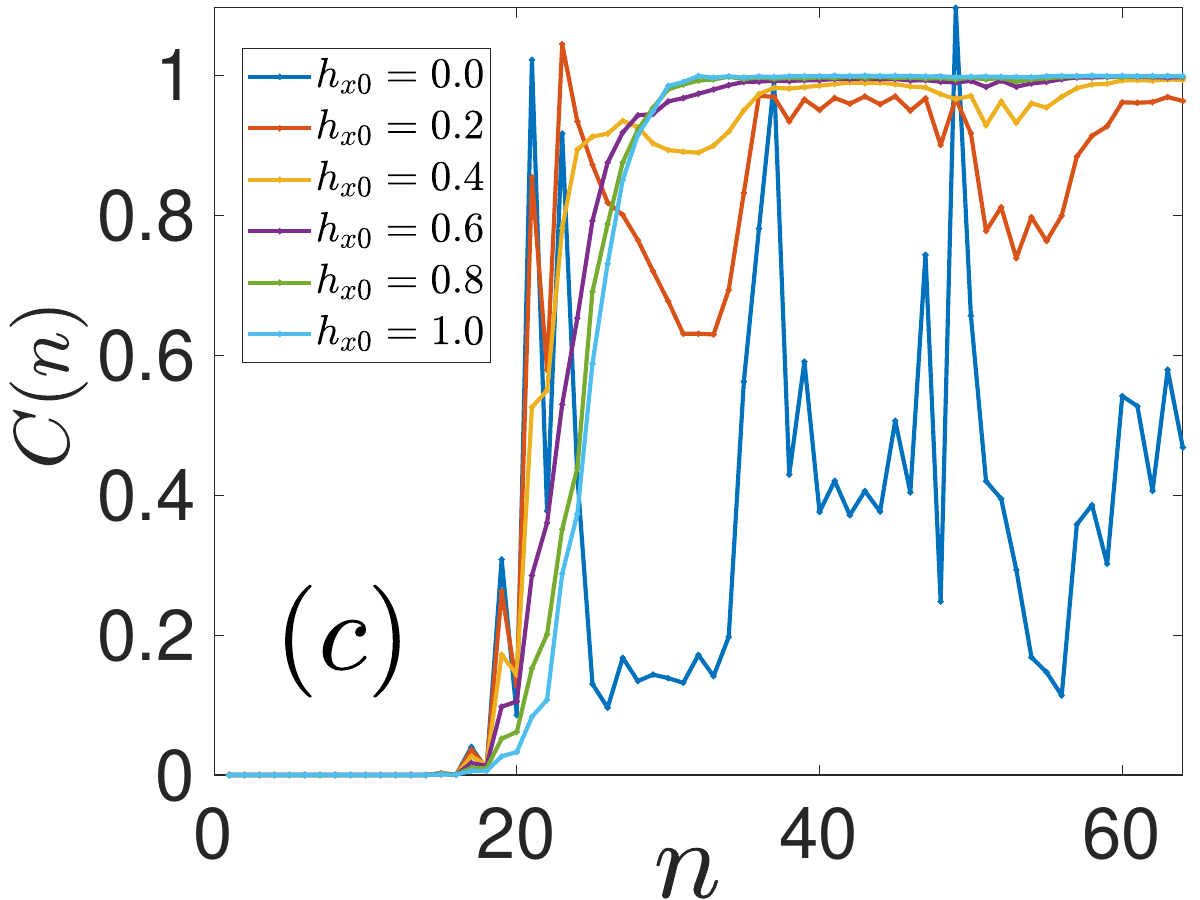}
       \caption{$C(n)$ ($\hat W=\hat S_1^z$, and $\hat V=\hat S_{N/2}^z$) vs $n$ for different value of  $h_{x0}$  lies in between $0$ to $1$ differing by $0.1$ for nonintegrable $\mathcal{\hat U}_{xp}$  system at period $(a)$ $\tau=\pi/16$, $(b)$ $\tau=\pi/6$, and $(c)$ $\tau=\pi/4$. Other parameters are: $J=1$, $h_{z0}=4$, $t_{\rm max}=16\pi$, $\alpha=\pi/2 t_{\rm max}$, and system size $N=12$. Open boundary conditions are considered. }
       \label{sz_OTOC_sat} 
\end{figure*}
\par
Similar to the main text case, in this case, OTOC displays the power-law growth $C_N(n)\sim n^b$ in both integrable and nonintegrable systems [Fig.~\ref{quench_block_hx_rand5_nint_p0_N18}($a$, $b$)]. The exponent of this power law is notably high ($b\approx11$). This observation resonates with prior research, confirming the characteristic behavior of integrable and nonintegrable systems \cite{shukla2022characteristic,shukla2022out,lin2018out}. 
\par
As we carried out our calculations while varying the system size $N$, we encountered an intriguing consistency: the growth of the OTOC remained remarkably stable. Within the dynamics regime, these growth patterns consistently remained parallel to each other across different system sizes [Fig.~\ref{quench_block_hx_rand5_nint_p0_N18}($c$)].  
\par
In both integrable and nonintegrable systems, the dynamic region of the OTOC demonstrates power-law growth. However, we find that the dynamic region is unable to effectively discriminate between integrable and chaotic systems. As a result, we shifted our focus to the saturation behavior of the OTOC. In the context of integrable spin systems, the saturation behavior of OTOCs exhibits a unique characteristic—oscillation. Specifically, the OTOCs display an asymmetric rise and fall within the saturation regime. Conversely, the nonintegrable spin systems paint a different picture. Here, the saturation behavior of OTOCs is marked by exact saturation. Unlike the oscillatory behavior seen in integrable systems, chaotic systems tend to settle into a fixed state at saturation, reflecting the chaotic nature of their dynamics. The intriguing aspect of our investigation lies in the consistency of this behavior across different periods. Whether we examine the systems at one, two, or even three periods $\tau=\pi/16$ [Fig.~\ref{OTOC_sat}($a$)], $\tau=\pi/6$ [Fig.~\ref{OTOC_sat}($b$)], and $\tau=\pi/4$ [Fig.~\ref{OTOC_sat}($c$)],  the behavior remains remarkably similar. In our observations, the value of $C_N(n)$ falls within the range of $0$ to $2$ due to the bounded nature of the asymptotic value of $C_4(n)/C(\infty)$, which remains between $-1$ and $1$ during the post-scrambling phase. Furthermore, the asymptotic value of $C_2(n)/C(\infty)$ converges to $1$.
\par
We calculate LSS for both integrable and nonintegrable time-dependent spin systems by fixing a period of $\tau=\pi/4$. The integrable realm presents us with a consistent observation: Poisson-type behavior in spectral statistics across all numbers of kicks. This consistent pattern reflects an integrable characteristic of the system throughout the time evolution of the system [Fig.~\ref{periodic_NNSD}($a-f$)]. However, the nonintegrable domain displays a different picture. Initially, the spectral statistics mimic Poisson-type behavior for a certain number of initial kicks. This behavior speaks to the presence of localized modes and partial regularity in the evolution of the system. However, after a distinct number of kicks, a shift occurs to Wigner-Dyson-type behavior, signifying the emergence of chaotic and globally randomized dynamics [Fig.~\ref{periodic_NNSD}($g-l$)]. This transition indicates that the system undergoes a transition to chaos after a certain number of kicks. Contrary to the behavior observed in the spin system with a linear field, the system displays a different pattern in the LSS in the spin system with a periodic field. In the linear field case, the system transitions to Wigner-Dyson behavior in the early stages of the evolution, occurring after just a few kicks. This is because the longitudinal field is fixed, while the transverse field starts with an initial small value $h_{x0}$. This setup initiates a competition between the longitudinal and transverse fields in the initial time regimes. In contrast, with a periodic field, the system shows Wigner-Dyson behavior after some additional kicks. Here, the longitudinal field starts from zero, and the transverse field starts from a high value of $h_x=4$. Initially, the transverse field dominates over the longitudinal field. However, as time progresses, the transverse field decreases due to its cosine form, while the longitudinal field increases due to its sine form. After some kicks, the effect of the longitudinal field becomes significant.
\par
Subsequently, we extended our study to consider localized Pauli spin observables in the x-direction at positions $1$ and $N/2$, {\it i.e.,} $\hat W = \hat \sigma_1^x$ and $\hat V = \hat \sigma_{N/2}^x$. In calculating the OTOC for a localized observable, we focus on the four-point correlation because the two-point correlation can be simplified to the identity operator due to the unitarity of Pauli observables. Consequently, in the limit as $n$ approaches infinity, a four-point correlation tends toward the identity \cite{shukla2022out}. Therefore, when calculating $C(n)$, it is not necessary to divide by $C(\infty)$. Remarkably, we found consistent saturation behavior of the OTOCs similar to what we observed with the nonlocalized spin observables. In the integrable case, the OTOCs exhibited oscillating saturation behavior, while in the chaotic case, we again observed exact saturation behavior at different time periods: $\tau = \pi/16$, $\tau = \pi/6$, and $\tau = \pi/4$ [Fig.~\ref{sx_OTOC_sat} ($a-c$)].

\section{Appendix: OTOCs using nonlocalized  and localized Pauli observables with $z$-direction alignment in time dependent Ising spin system}
\label{sat_local}
The behavior of OTOC depends on the choice of observables, it is crucial to analyze OTOC with different observables to establish the saturation region as a universal discriminator. To this end, we define nonlocalized observables aligned in the z-direction as $\hat W =\frac{2}{N} \sum_{i=1}^{N/2}\hat S_i^z$ and $\hat V = \frac{2}{N}\sum_{i=N/2+1}^{N}\hat S_i^z$. Our investigations cover both integrable and nonintegrable systems under the influence of the operator $\hat{\mathcal{U}}_{xp}$. For the OTOC calculations, we consider three periods: $\tau = \pi/16$, $\tau = \pi/6$, and $\tau = \pi/4$. In all cases, we observe the oscillatory behavior of the OTOC in the integrable case, while in the chaotic case, we observe exact saturation behavior [see Fig.\ref{Block_z_OTOC_sat} ($a-c$)]. When the period is short, the saturation regime of the OTOC shows an increasing trend [see Fig.\ref{Block_z_OTOC_sat} (a)]. However, compared to the integrable case, we observe a non-oscillating behavior in the chaotic case. As the period increases, for large amplitudes of the longitudinal field (around $1$), we observe exact saturation [see Fig.\ref{Block_z_OTOC_sat} (b,c)]. This suggests that when the longitudinal field is comparable to the transverse field, the system undergoes in chaotic regime.
\par
Finally, we considered localized Pauli spin observables in the z-direction at positions $1$ and $N/2$, {\it i.e.,} $\hat W = \hat \sigma_1^z$ and $\hat V = \hat \sigma_{N/2}^z$. Much like the previous cases, we found consistent saturation behavior of the OTOCs, resembling the behavior seen with the nonlocalized observables. In the integrable case, the OTOCs displayed oscillating saturation behavior, while in the chaotic case, we again observed exact saturation behavior at different time periods: $\tau = \pi/16$, $\tau = \pi/6$, and $\tau = \pi/4$ [Fig.~\ref{sz_OTOC_sat} ($a-c$)].

\vspace{0.5cm}
{\bf Author contribution statement} \\
\noindent
RKS, GRM, SA, and SKM conceived the idea, developed the theory, and conducted the analysis. RKS and GRM performed numerical calculations, interpreted the results, and contributed to the final manuscript.

\vspace{0.5cm}
{\bf Data Availability Statement} This manuscript has no associated data or the data will not be deposited. [Authors’ comment: All the data is already displayed in the figures of the manuscript.]

\bibliographystyle{apsrev4-1}
\bibliography{Quench_scramb}
\end{document}